\documentclass[journal=jpcafh,manuscript=article]{achemso}

\usepackage[version=3]{mhchem} 
\usepackage{amsmath,amssymb,amsfonts}
\usepackage{xcolor} 
\usepackage{epsfig}
\usepackage{amssymb}
\usepackage{subfig}
\usepackage{graphicx}
\usepackage{epsfig}
\usepackage{longtable}
\usepackage{bm}
\usepackage{array}
\usepackage{lscape}
\usepackage{amsmath}

\usepackage{subfig}
\usepackage{subfloat}

\author{Aleksander P. Wo{\'z}niak}
\affiliation[University of Warsaw]{Faculty of Chemistry, University of Warsaw, Pasteura 1, 02-093 Warsaw, Poland}
\email{ap.wozniak@uw.edu.pl}
\author{Ludwik Adamowicz}
\affiliation[University of Arizona]{Department of Chemistry and Biochemistry, University of Arizona,  
1306 E University Blvd,
Tucson, Arizona 85721-0041, USA}
\author{Thomas Bondo Pedersen}
\author{Simen Kvaal}
\affiliation[University of Oslo]{Hylleraas Centre for Quantum Molecular Sciences, Department of Chemistry, University of Oslo, 
P.O. Box 1033 Blindern, N-0315 Oslo, Norway}

\title[Gaussians for Quantum Dynamics]{Gaussians for Electronic and Rovibrational Quantum Dynamics}

\begin{document}

\begin{tocentry}

\includegraphics[width=1.\linewidth]{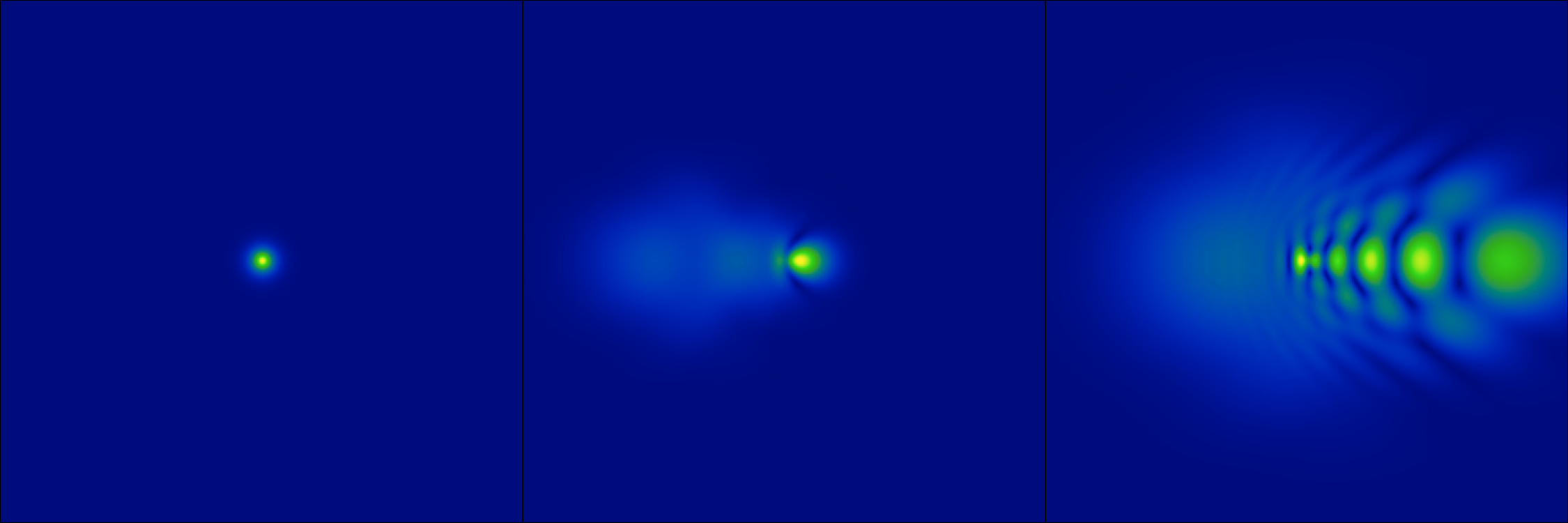}

\end{tocentry}

\begin{abstract}    
The assumptions underpinning the adiabatic Born-Oppenheimer (BO) approximation are
broken for molecules interacting with attosecond laser pulses, which generate
complicated coupled electronic-nuclear wave packets that generally will have
components of electronic and dissociation continua as well as bound-state contributions.
The conceptually most
straightforward way to overcome this challenge is to treat the
electronic and nuclear degrees of freedom on equal quantum-mechanical
footing by \emph{not} invoking the BO approximation at all.
Explicitly correlated Gaussian (ECG) basis functions have proved successful
for non-BO calculations of stationary molecular states and energies, reproducing
rovibrational absorption spectra with very high accuracy.
In this paper, we present a proof-of-principle study of the ability of fully flexible ECGs (FFECGs)
to capture the intricate electronic and rovibrational dynamics generated by short, high-intensity
laser pulses. By fitting linear combinations of FFECGs to accurate wave function histories
obtained on a large real-space grid for a regularized 2D model of the hydrogen atom and for the 
2D Morse potential we demonstrate that FFECGs provide a very compact description of 
laser-driven electronic and rovibrational dynamics.
\end{abstract}

\section{Introduction} \label{introduction}

With the advent of new technology for manipulating atoms and
molecules with intense ultrashort (atto- and femtosecond)
laser pulses, there is an urgent need for
further development of accurate and 
reliable quantum-dynamics (QD)
tools for simulations
of events involved in such manipulations.
Such simulations are needed both to guide the design
of experiments and to ensure correct interpretation of observations.
Atomic and molecular QD simulations involving
interaction of these systems with ultrashort intense laser pulses
can be carried out by integrating the time-dependent 
Schr\"{o}dinger
equation (TDSE) on a real-space grid or by using an expansion 
of the wave function of the system in terms of 
basis functions whose
linear and non-linear parameters are adjusted along with the number of
basis functions during the propagation.
The focus of this work is an analysis of
the necessary features such basis functions must possess to 
be effective for simulating laser-induced dynamics of an atomic or a molecular system.

The dynamics induced by the interaction of 
intense electromagnetic radiation with nuclei and 
electrons of a molecule necessitate an accurate
account of the coupling of these two types of particles.
The broad intensity distribution in the frequency domain of ultrashort laser pulses
implies that a large number of bound and continuum states
are involved in the dynamics, including rovibrational, electronic,
and collective states where the motion of both nuclei and electrons
are simultaneously excited.
To describe such an intricate
situation, the separation of the nuclear and electronic 
motions---a hallmark of the adiabatic Born-Oppenheimer (BO)
approximation\cite{Born1927,Born1954}---must not be assumed and,
ideally, all particles forming the system should be treated on an equal footing.
Such an approach is tested in the present work 
using two simple 2D models.
Also, an extension of the approach to 
simulate the laser-induced dynamics 
of attosecond atomic and molecular
events involved in attosecond experiments
\cite{Corkum2007,Nisoli2017,Borrego-Varillas2022}
is discussed.

The coupled nuclear-electronic motion is highly
correlated,
as the electrons, particularly the core electrons,
generally follow the nuclei very closely and the nuclei 
stay apart from each other 
due to their strong Coulomb repulsion and large masses.
The situation is markedly different for the
electrons whose wave functions,
due to much lower masses, more 
significantly
overlap. To best describe these effects
using a basis-set approach,
one needs to expand the
wave function in terms of functions that explicitly
depend on nucleus-nucleus, nucleus-electron, and 
electron-electron distances, i.e.,
the explicitly-correlated functions (ECFs). In the first part of this work, we 
review the ECFs used in the atomic and molecular
calculations of stationary bound states and
we discuss the features of these functions
that may make them particularly useful in
QM calculations of atomic and molecular systems.
We particularly
focus on Gaussian ECFs (ECGs),
as these are the most popular functions 
used in non-BO atomic and molecular 
calculations
\cite{Bubin2013,Kozlowski19932007,Mitroy2013,Bubin2005377,Matyus2019590,Simmen2014,Muolo2018_JCP,Muolo2018,Matyus2012,Muolo2020,Johnson2016,Valeev2006,Ten-no2003152,Takatsuka2003859,Varga2019}.
In the second part, two-dimensional
time-propagation calculations are performed 
for two model systems
involving Coulomb and Morse potentials using a grid 
approach. Next, the time-dependent
grid wave functions are fitted with 
ECGs that are chosen to 
best represent the key features that
appear in the wave function due to the interaction of the 
system with ultra-short intense laser pulses.

Single-particle Gaussians have been extensively used as basis functions in both electronic-structure theory \cite{MEST} and
vibrational dynamics \cite{Heller1975,Heller1976,Heller1981}.
In electronic-structure theory the Gaussians are real-valued functions centered 
at the atomic nuclei and contracted to form atomic orbitals which, in turn, 
form a non-orthogonal basis for the expansion of molecular orbitals---see,
e.g., Ref.~\citenum{MEST} for a detailed account.
Such Gaussians have also been used for the study of many-electron
dynamics \cite{Goings2018,Li2020,ofstad_time-dependent_2023}, although
important highly nonlinear phenomena such as ionization processes and high
harmonic generation cannot be properly accounted for.
The latter can to a certain extent be ameliorated by augmenting the standard
basis with Gaussians fitted to continuum (Bessel, Coulomb, or Slater-type)
functions; see the recent review by \citeauthor{Coccia2022}
and references therein for more details \cite{Coccia2022}.

The core idea of 
\citeauthor{Heller1975}'s approach \cite{Heller1975,Heller1976,Heller1981} 
to vibrational dynamics
is to use complex-valued Gaussians (Gaussian wave packets),
which are exact solutions for harmonic potentials. Unlike in electronic-structure
theory, the Gaussian parameters are now time-dependent variational
parameters. For anharmonic potentials,
however, the equations of motion for the Gaussian parameters quickly
become ill-conditioned and one resorts to locally harmonic approximations
of the potential, which is a reasonable approach as long as the wave packet is
sufficiently localized---see, e.g., Ref.~\citenum{Vanicek2020} for a
recent review of the so-called thawed Gaussian approach.
Complex-valued Gaussians have also been used in the context of the
multiconfigurational time-dependent Hartree (MCTDH)
method \cite{Meyer_etal2009,worth_full_2003,worth_novel_2004,Burghardt_etal2008}. In all cases, however, the ill-conditioned equations of motion is 
a serious obstacle. In this work, we investigate the ability of complex-valued
Gaussians to represent complicated dynamics of electrons and nuclei by 
fitting to accurate grid-based solutions of the time-dependent Schrödinger
equation.

\section{Non-BO Hamiltonian} \label{hamiltonian}

The total non-relativistic all-particle non-BO molecular Hamiltonian
describing the interaction of a neutral molecule 
with a uniform, time-dependent electric field defining
the $x$-axis of a laboratory-fixed coordinate frame
can be rigorously separated into a center-of-mass (COM)
kinetic-energy operator and the internal Hamiltonian
\cite{Thomas1970,Bubin2013,Mitroy2013},
$\hat{H}(t)$. The separation is accomplished by 
transforming the total Hamiltonian from
Cartesian laboratory coordinates,
$\mathbf{R}_i$, $i = 1, \dots, N$ ($N$ is the total
number of particles in the molecule)
to a new Cartesian coordinate
system, parallel to the laboratory frame, where the first three coordinates are the 
coordinates of the COM and the remaining coordinates 
are internal coordinates.
The origin of the internal frame is chosen at a reference
particle, typically the heaviest nucleus, which is taken to be particle number $1$ such that
$\boldsymbol{r}_i = \boldsymbol{R}_{i+1}-\boldsymbol{R}_1$ for $i=1,\dots,n$ with $n=N-1$.
The internal Hamiltonian then takes the following form (using atomic units throughout)\cite{Bubin2013}:
\begin{equation}
\label{internal}
    \hat{H}(t) = \sum_{i=1}^n \left(
                       -\frac{1}{2\mu_i} \nabla_{{\bf r}_i}^2 + \frac{q_0 q_i}{r_i}
                 \right)
               + \sum_{i<j}^n \left(
                       \frac{q_i q_j}{r_{ij}} + \frac{1}{M_1} \nabla_{{\bf r}_i}^\prime \nabla _{{\bf r}_j}
                 \right)
               - \mathcal{F}(t) \sum_{i=1}^{n} q_i x_i,
\end{equation}
where $\mathcal{F}(t)$ is the time-dependent electric-field strength,
$M_1$ is the mass of the reference particle (particle $1$),
$q_i = Q_{i+1}$ ($i=0,\dots,n$), 
$\mu _i = M_1 M_{i+1}/(M_1 + M_{i+1})$ ($i = 1, \dots, n$) with
$Q_i$ and $M_i$, $i = 1, \dots, N$, the charge and mass
of particle $i$,
$r_{ij}=| {\bf r}_i-{\bf r}_j| =
| {\bf R}_{i+1}-{\bf R}_{j+1}|$, 
$ r_i = | {\bf r}_i |$, and the prime denotes vector/matrix
transposition.
The Hamiltonian \eqref{internal} represents $n$
interacting particles with masses equal to the reduced masses
moving in the central Coulomb potential of the reference particle.
We refer to these
particles as pseudo-particles because, while they have the same charges
as the original particles, their
masses are different.
For $\mathcal{F}(t)=0$ the Hamiltonian \eqref{internal}
is fully symmetric (isotropic or atom-like) 
with respect to all rotations around the center of the
internal coordinate system and its eigenfunctions
transform as irreducible representations 
of the fully symmetric
group of rotations.
When $\mathcal{F}(t) \neq 0$, however,
the symmetry is reduced to cylindrical about the field direction, here the $x$-axis.

For a diatomic system, after separation of the center of mass motion, the internal Hamiltonian used in the non-BO time-evolution calculations represents a motion of the second nucleus and the electrons (with their masses replaced by the respective reduced masses) in the field of the charge of the first nucleus (the reference nucleus) located in the center of the internal coordinate system. The potential acting on the second nucleus that results from the interaction of this nucleus with the charge of the reference nucleus and the electrons can effectively be represented by a Morse-like potential. An important effect that also determines the electronic-nuclear dynamics of the system is the electrostatic attractive interaction of each of the electrons with the reference nucleus located at the center of the coordinate system. Thus, the Coulombic and Morse interactions investigated in this work are central to understanding the molecular dynamics. The interactions which are present in the internal Hamiltonian, but are not investigated in this work, are two particle interactions involving the second nucleus and the electrons, and the inter-electron interactions. To represent these types of interactions, models involving more than two dimensions would be needed and, thus, they are not investigated in this work. However, based on our ECG stationary-state calculations, we expect that FFECGs should perform very well in describing these interactions.

\section{ECGs used in very accurate non-BO calculations of stationary states of small atoms and molecules} \label{ecgs}

To achieve high accuracy needed in the computations we will
use an approach that is both adaptive in space and time.
Mesh-free complex explicitly-correlated 
Gaussian (CECG) functions, that are free to warp and
roam in space, will be the main tool. 
The Adamowicz group has used 
various types of ECGs and CECGs in
very accurate non-BO atomic and molecular calculations
of stationary bound states for over two decades.
Various forms of explicitly correlated
all-particle Gaussian functions (ECGs) with 
real and/or complex 
non-linear parameters have been used in non-BO calculations
\cite{Bubin2013,Kozlowski19932007,Mitroy2013,Bubin2005377}.
The simplest ECG with real non-linear parameters used
to calculate an $S$ state of an $n$-electron atom ($S$-ECG) is:
\begin{equation}
\phi({\mathbf r}) = 
\exp[-{\mathbf r}^{\prime} {\mathbf A} {\mathbf r}],
\end{equation}
where $\mathbf r$ is vector of $3n$ internal Cartesian 
coordinates of the electrons 
and $\mathbf A$ is a $3n \times 3n$ real 
symmetric positive-definite matrix of the non-linear
parameters. ${\mathbf A}$ has the following block structure:
$\mathbf{A} = A \otimes I_3$, 
where $A$ is a $n \times n$ real dense symmetric 
positive-definite matrix and 
$I_3$ is a $3\times 3$ identity matrix, while 
symbol $\otimes$ denotes the Kronecker product. 
Such representation of matrix 
${\mathbf A}$ ensures that the exponential part of the 
basis function is invariant
with respect to 3D rotations.

$S$-ECGs can alternatively be represented as: 
\begin{eqnarray}
\phi({\mathbf r}) =
\exp \left[
-\alpha_1 {\mathbf r}_1^2 -\alpha_1 {\mathbf r}_2^2 - 
\dots -\alpha_n {\mathbf r}_n^2 \right]\times \\ \nonumber
\exp\left[-\beta_{12} {\mathbf r}_{12}^2 - \beta_{13} 
{\mathbf r}_{13}^2 - \dots - \beta_{(n-1)n} 
{\mathbf r}_{(n-1)n}^2\right],
\end{eqnarray}
where the first factor is a product
of $n$ orbitals and the second factor
is a product of $n(n+1)/2$ 
pair functions explicitly dependent on
the squares 
of all inter-electron distances, ${\mathbf r}_{ij}^2$. 
The methods allow for very 
accurate calculations of the spectra of small 
atoms and molecules when the leading relativistic and
QED corrections are also included in the calculations.

The non-BO ECG calculations for
$S$, $P$, $D$, and $F$ 
states of atomic systems with 2-5 electrons
\cite{Stanke2022,Hornyak2021,Hornyak2020,Stanke2019,Hornyak2019,Stanke2018,Bubin2017,Bubin2017a,Stanke2022a,Nasiri2022,Stanke2023263,Nasiri2022a,Stanke2023}  
are among the most accurate in
the literature. 
As the ECGs explicitly depend on 
the distances between the particles (electrons and nuclei), 
they very efficiently 
represent the coupled nucleus-electron 
motions and allow very accurate accounting of the
inter-particle correlation effects. These 
effects are indispensable in non-BO calculations, 
because, as mentioned, the
Coulomb interactions make particles with alike 
charges avoid each other and particles with
opposite charges to follow each other. 
This effect can also be very 
effectively described with the ECGs.

A challenge in non-BO ECG 
calculations of stationary ground and excited states
of small atoms and molecules
is to accurately describe radial and angular 
oscillations of the non-BO wave functions
of highly excited states.
Three types of ECGs have been used to 
describe these features. 
These are: \\
{\bf (a)} molecular ECGs with pre-exponential multipliers 
in the form
of powers of the inter-nuclear distances.
The functions are called "power" ECGs (PECGs)
and they have the following form:
\begin{equation}
\phi({\mathbf r}) = 
\prod_i r_i^{m_i} \prod_{i>j} r_{ij}^{m_{ij}}
\exp[-{\mathbf r}^{\prime} {\mathbf A} {\mathbf r}],
\label{pecg}
\end{equation}
where $m_i$ and $m_{ij}$ are even non-negative 
integers, and ${\mathbf A}$, as defined before, is a real symmetric 
positive-definite $3n\times 3n$ matrix.
PECGs have been used in molecular non-BO calculations 
for small molecules
\cite{Scheu20013393,Kinghorn20004203,Kinghorn19992541,Kinghorn19978760,Bubin2009,Nasiri2022,Bubin2005}; \\
{\bf (b)} complex single-center ECGs.
The works that are particularly relevant to this project
concern implementation of algorithms for performing 
very accurate calculations on bound states of
small molecules that employ complex ECGs (CECGs):
\begin{equation}
\exp[-{\mathbf r}^{\prime}({\mathbf A} + {\rm i} {\mathbf B})
{\mathbf r}],
\label{cecg}
\end{equation}
where ${\mathbf A}$, as defined before, is a real
symmetric matrix and ${\mathbf B}$ is also
a real symmetric matrix with the same structure as ${\mathbf A}$
(i.e. ${\mathbf B} = {\mathbf b} \otimes {\mathbf I}_3$,
where ${\mathbf b}$ is a real dense symmetric $n \times n$ matrix). 
It was shown that CECGs can very effectively 
describe high-frequency radial 
oscillations of the wave function
of highly vibrationally excited states. 
The angular oscillations can be described by adding 
Cartesian spherical harmonics as 
pre-exponential multipliers
to the Gaussians.
CECGs have been used in non-BO calculations 
of molecular 
$\Sigma$, $\Pi$, and $\Delta$ bound 
rovibrational states
\cite{Bubin2020204102,Chavez2019147,Bubin2017b,Bubin2016122}
It has been shown that CECGs are equally, if not more, 
efficient as PECGs in describing radially and angularly
oscillating wave functions of states
located near the dissociation threshold; \\
{\bf (c)} real ECGs with shifted centers (SECGs) 
of the form:
\begin{equation}
\phi({\mathbf r}) = \exp[-({\mathbf r} - {\mathbf q})^{\prime} 
{\mathbf A} 
({\mathbf r} - {\mathbf q})],
\label{SECG}
\end{equation}
where ${\mathbf q}$ is $3n$ real vector of the 
Gaussian shifts and
${\mathbf A}$, as defined before, is a real symmetric matrix of the 
non-linear parameters.
SECGs have been used 
in non-BO calculations of some small diatomic and triatomic
molecules and in non-BO calculations of the dipole moments,
polarizabilities, and hyperpolarizabilities of isotopologues
of the H$_2$ molecule.
\cite{Cafiero2004,Cafiero20072679,Cafiero2005,Cafiero2003113,Cafiero20025557,Cafiero2002,Adamowicz2020a,Adamowicz2018}
Including real shifts in the Gaussians enables 
to describe radial and angular
polarization of the wave function. 
These types of deformations 
can be also described by linear combinations of 
spherical-harmonics factors, though the shifts may 
be a more effective way for the task.


The purpose of this work is to develop and test 
an ECG basis set to be employed in 
quantum-dynamics simulations of molecular systems
exposed to an ultrashort laser pulse within the
semiclassical electric-dipole approximation.
The proposed basis is tested by fitting
wave packets obtained as solutions of the time-dependent 
Schr\"{o}dinger equation with a grid-based method 
for two two-dimensional (2D) model systems. The models
represent two main features that need to be described
in a QD simulation of a molecule, i.e.,
the electrostatic interaction represented by a Coulombic 
potential and
the rovibrational interaction represented 
for a diatomic molecule by a Morse potential. 
In the next section, before the ECG basis functions
for QD molecular simulations are introduced, the grid-based
calculations of the two models are described
and discussed. 

\section{2D Model calculations using a grid approach} \label{calculations}

\begin{figure}
  \centering
  \includegraphics[width=0.49\textwidth]{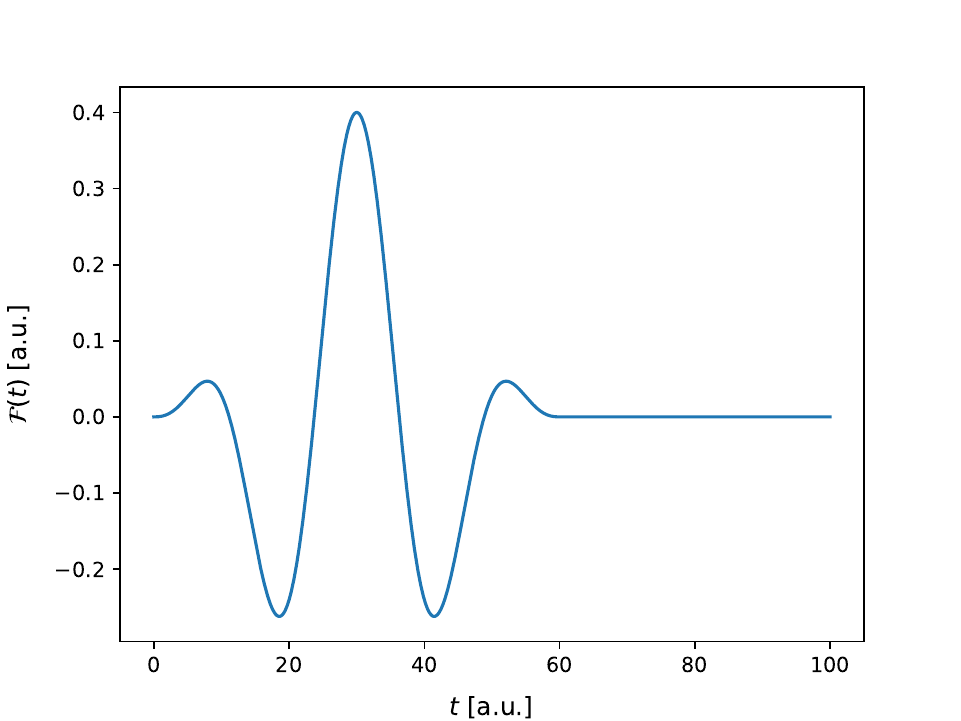}%
  \includegraphics[width=0.49\textwidth]{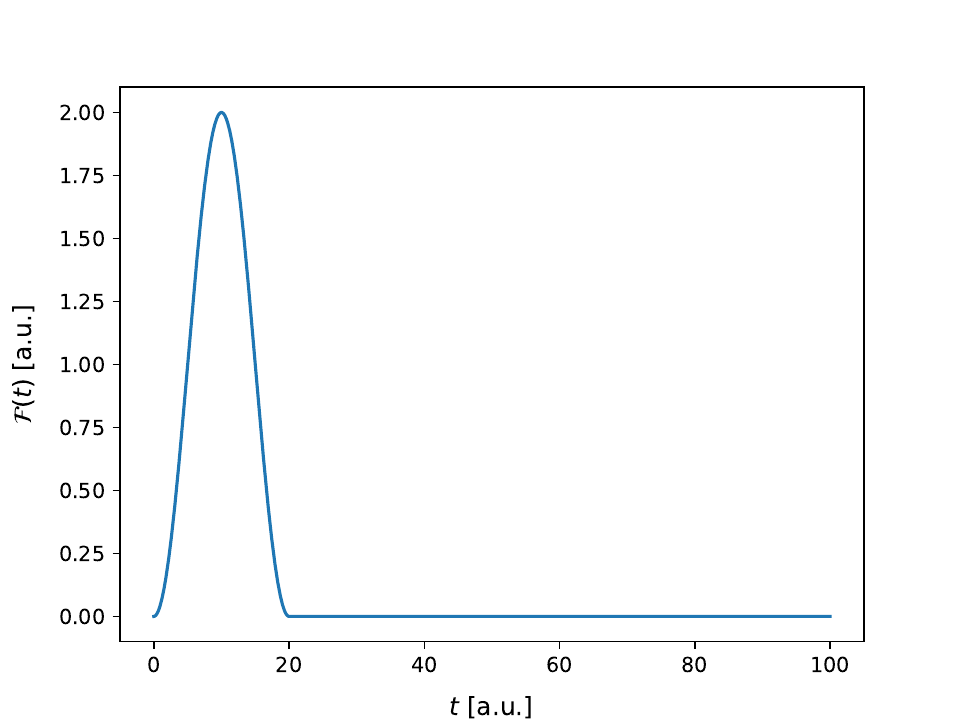}
  \caption{ Shape of the laser pulse used in the simulation of the Coulomb model (left) and of the Morse model (right) }
  \label{fig:pulse}
\end{figure}

Our ultimate goal is to solve the 
TDSE, ${\rm i} \dot{\psi}(t) - \hat{H}(t)\psi(t) = 0$ 
for the non-BO internal Hamiltonian (\ref{internal})
representing a molecule interacting with a short
intense laser pulse. For a diatomic molecule,
there are two major interactions that need to be
described. The first is the repulsive interaction of the 
pseudo-nucleus with the charge of the reference nucleus
located in the center of the coordinate system,
and the second is the attractive Coulomb interaction
of a pseudo-electron with the charge of the reference
nucleus. Due to the screening effect
of the former interaction by the electrons, the 
interaction potential is not Coulombic but is 
more appropriately represented 
by a Morse-type potential.
Thus, at the very minimum, in selecting an ECG basis
set for solving the non-BO TDSE, one should
verify if the chosen basis is capable of describing
the laser-induced dynamics of a single 
particle interacting with 
the charge of the reference particle with 
the Morse and Coulomb potentials. 
For the verification we use 
an elementary 2D model Hamiltonian of the form:
\begin{equation}
    \hat{H}(t) = -\frac{1}{2\mu}
    \left (\frac{\partial^2}{\partial x^2}
    +      \frac{\partial^2}{\partial y^2} \right ) 
    + V(x,y) - q x \mathcal{F}(t).
\end{equation}
For the electron we use the soft Coulomb potential,
$V(x,y) = -(x^2 + y^2 + 1/2)^{-1/2}$, which 
mimics the nuclear potential of a hydrogen-like atom, and
the Morse potential is given by
$V(x,y) = D_e 
\left[ 1 - \exp (-\alpha ((x^2+y^2)^{1/2} - r_e)) \right]^2$, with $D_e = 0.17449$, $r_e = 1.4011$, and $\alpha = 1.4556$.
The charge and (reduced) mass are set to $q=-1$, $\mu=1$ for the
Coulomb model, and $q=1$, $\mu=1605.587$ for the Morse model.
The electric-field strength
is nonzero only in the time 
interval $t_0 < t < t_1$, where it is equal to:
\begin{equation}
    \mathcal{F}(t) = \mathcal{E}_0\sin^2 \left(\pi \frac{t-t_0}{t_1 - t_0}\right)\cos(\omega (t-\bar{t})), \qquad \bar{t} = \frac{t_0 + t_1}{2},
\end{equation}
where $\mathcal{E}_0$ denotes the maximum electric field amplitude. 
In our calculations for both models we set $t_0 = 0$.
For the Coulomb model we set $\omega = 0.25$ a.u., $t_1 
= 60$ a.u. and $\mathcal{E}_0 = 0.4$ a.u., which corresponds to a laser pulse of wavelength $\approx 182$ nm, consisting of 2.5 optical cycle.
For the Morse model we set $\omega = 0.0$ a.u., $t_1 
= 20$ a.u. and $\mathcal{E}_0 = 2.0$ a.u., corresponding to a short, delta-like pulse (relative to the
time-scale of the nuclear motion).
Our laser pulse 
parameters are chosen not for their significance in 
relation to any experiment, but rather such that it 
generates complicated ionization and dissociation dynamics.
We consider the 
dynamics for times $0 \leq t \leq 100$ a.u. for the Coulomb model and $0 \leq t \leq 300$ a.u. for the Morse model, including periods 
of free evolution after the laser pulse. 
The laser pulses used in the QD simulations of
the Coulomb and Morse models are shown
in Fig.~\ref{fig:pulse}.

Highly accurate reference QD calculations
with the Coulomb and Morse potentials 
are performed by spatially discretizing 
the real $xy$ plane using a grid with $n_\text{grid}=1024$ 
equidistant points in the interval $[-L,L] = [-150,150]$ for the Coulomb model and $[-20,20]$ for the Morse model,
in both directions.
The kinetic-energy operator is approximated using
the standard Fast Fourier Transform (FFT),
which introduces
artificial periodic boundary conditions. These have 
negligible effect due to the large domain.
The time evolution can be approximated in a number of ways, but we 
choose the common second-order split-operator scheme\cite{feit_solution_1982} with time step $h = 0.01$ a.u. for the Coulomb model and $h = 0.05$ a.u. for the Morse model, respectively. This method has accuracy of order $\mathcal{O}(h^3)$ locally in time, and is sufficiently accurate for our purposes.
The calculations are initiated with the 
the corresponding ground-state wave functions
which are fully symmetric with respect to all 
rotations around the center of the internal 
coordinate system in the $xy$ plane.
The ground-state wave function is obtained using inverse iterations using the conjugate-gradient method for the solution of large sparse linear systems.
In the case of the Coulomb potential the ground-state
wave function approximates a 2D $1s$ orbital and,
in the case of the Morse potential, the ground-state
wave function has a "torus" shape and 
is practically zero at the coordinate 
center, peaks at $r_e$, and then again goes to zero
at larger distances.
In Fig.~\ref{fig:hydrogen_grid} and Fig.~\ref{fig:morse_grid}, 
some snapshots of the time evolution of the wave
functions are shown for the Coulomb and Morse 
simulations, respectively. 
For each case, the real and imaginary 
parts of the wave function, as well as the 
wave-function absolute value, are plotted.
In both cases the respective wave functions become
increasingly more complicated with many features, 
more deformed and
oscillatory, and more diffused. 

It is interesting to know to what extent the 
time-evolving wave packets for the two considered models
involve contributions from higher angular momenta. 
For both systems, at $t=0$, the initial 
wave packets, i.e. the corresponding wave functions, 
are fully rotationally symmetric, i.e., 
for both $\hat{L}_z \psi(0) = 0$, where $L_z = -\text{i}\frac{\partial} 
{\partial\theta}$ is the angular 
momentum operator in 2d and $\theta$ 
is the angular coordinate in planar polar coordinate
system. 
Moreover, $[\hat{H}(t),\hat{L}_z]=0$ before and after the 
pulse begins and after it ends. 
The non-vanishing of the commutator for $t_0 < t < t_1$ 
implies that $\psi(t)$ is not, in general, an eigenfunction of 
$\hat{L}_z$. 

\begin{figure}
    \centering
    \includegraphics{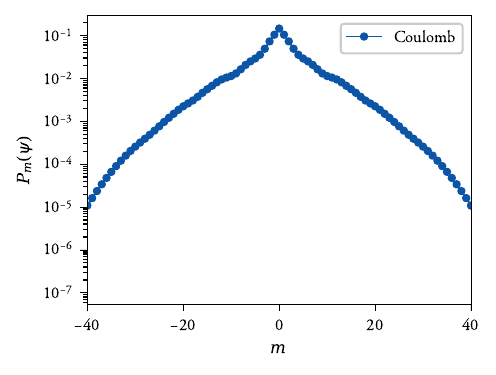}%
    \includegraphics{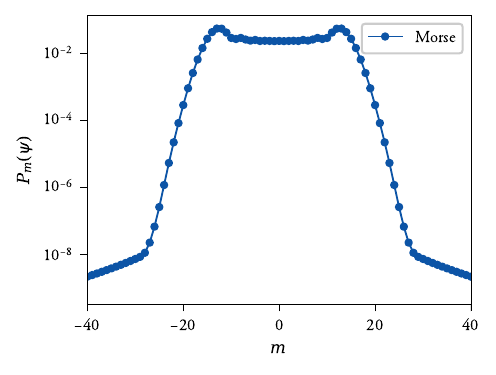}
    \caption{ Left: Angular momentum probability distribution of the Coulomb model wave function at $t=45$. Right: Same for the Morse model at $t= 20$ a.u. While the Morse model is bimodal with strong peaks around $|m|=13$, the Coulomb model is unimodal, with standard deviation $\sigma_m = 6.9$. The spectra are computed by first sampling the grid wave function at a polar coordinate grid using high-order spline interpolation, and then Fourier decomposing the result along the angular coordinate, resulting in $\psi(t) = \sum_{m} (2\pi)^{-1/2} e^{im\theta} f_m(r,t)$. The probabilities plotted are defined as $P_m(\psi(t)) = \int_0^\infty |f_m(r,t)|^2 r \; dr$, and are computed using numerical quadrature.}
    \label{fig:Lz-spec}
\end{figure}

In Figure~\ref{fig:Lz-spec}, the angular momentum probability distributions are shown for both model systems. Clearly, the pulse induces high angular momenta in the wave function.

\begin{figure} 
  \centering
  \subfloat{\includegraphics[width=0.9\textwidth]{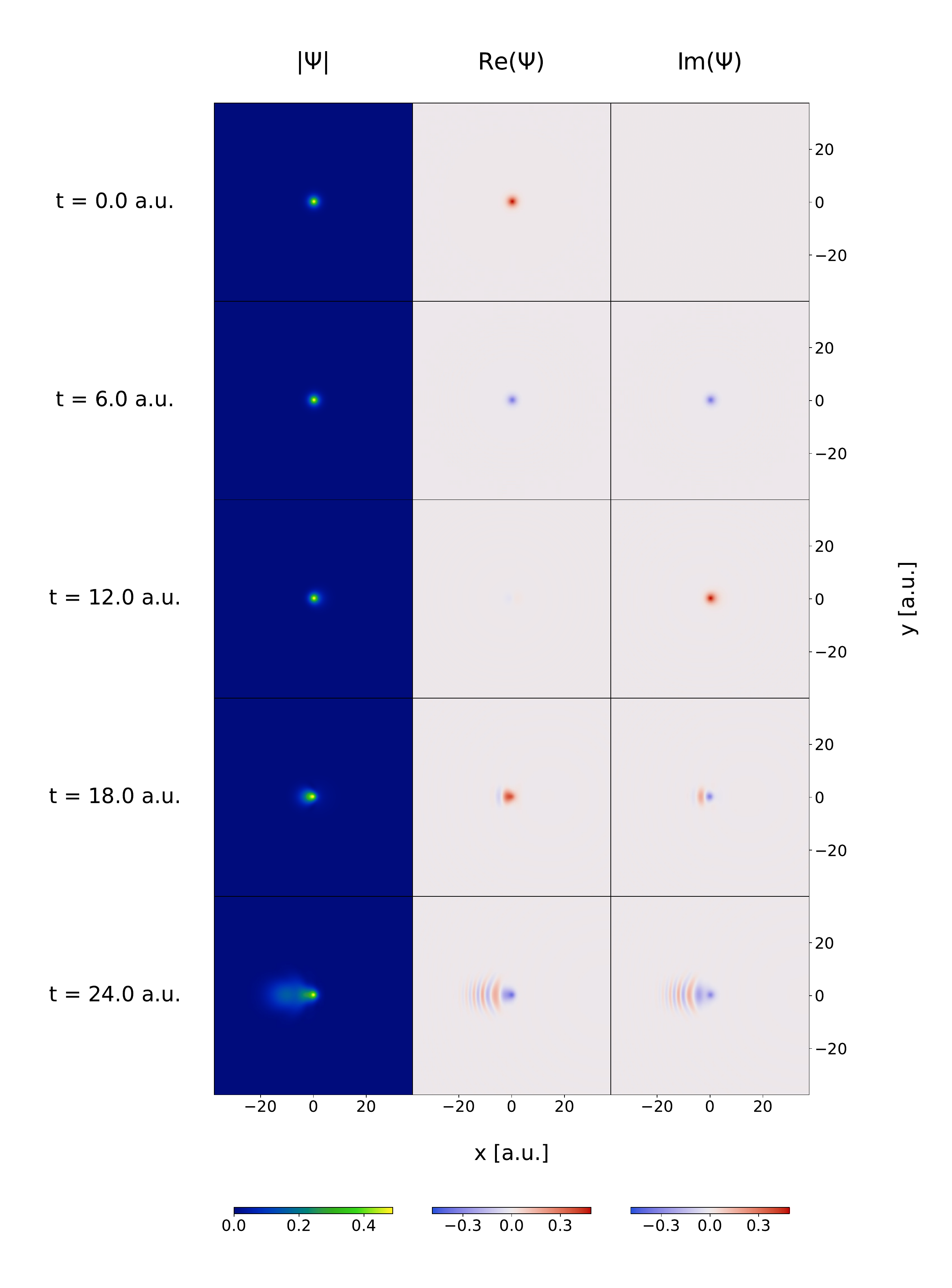}}
  \captionsetup{labelformat=empty}
  \caption{}
\end{figure}
\begin{figure}
\ContinuedFloat
  \subfloat{\includegraphics[width=0.9\textwidth]{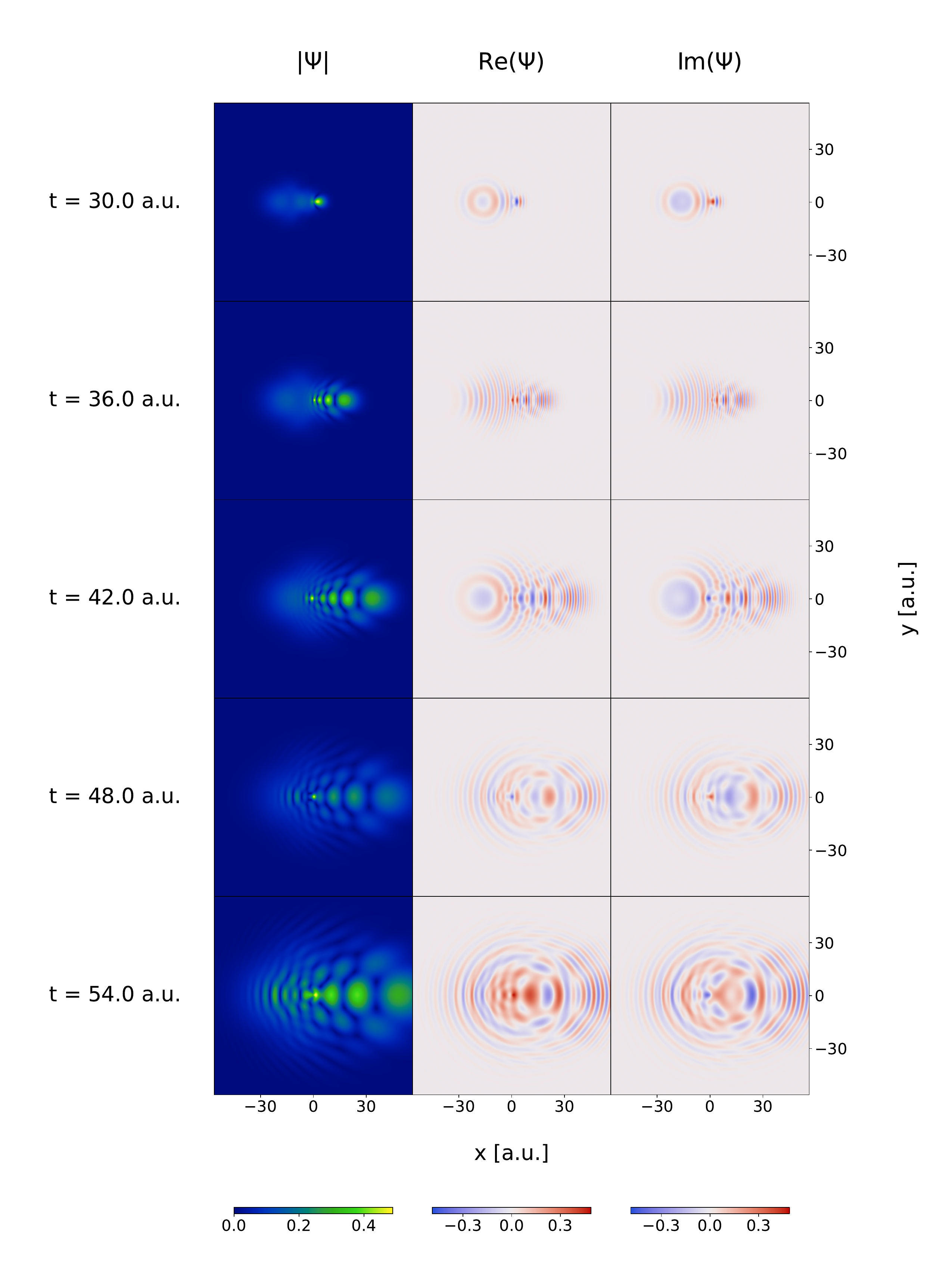}}
  \caption{ Snapshots illustrating the time evolution of the Coulomb model wave function during the grid-based simulation in the time interval from $t=0.0$ to $t=54.0$ a.u.
  As the time advances, the wave function becomes progressively more complicated, with nonlinear phase, amplitude oscillations, and localized features.
  To facilitate visualization, the wave function values have been rescaled so that the maximum value remains constant across all timeframes, matching the maximum value at $t=0$.}
\label{fig:hydrogen_grid}
\end{figure}

\begin{figure}
  \centering
  \subfloat{\includegraphics[width=0.9\textwidth]{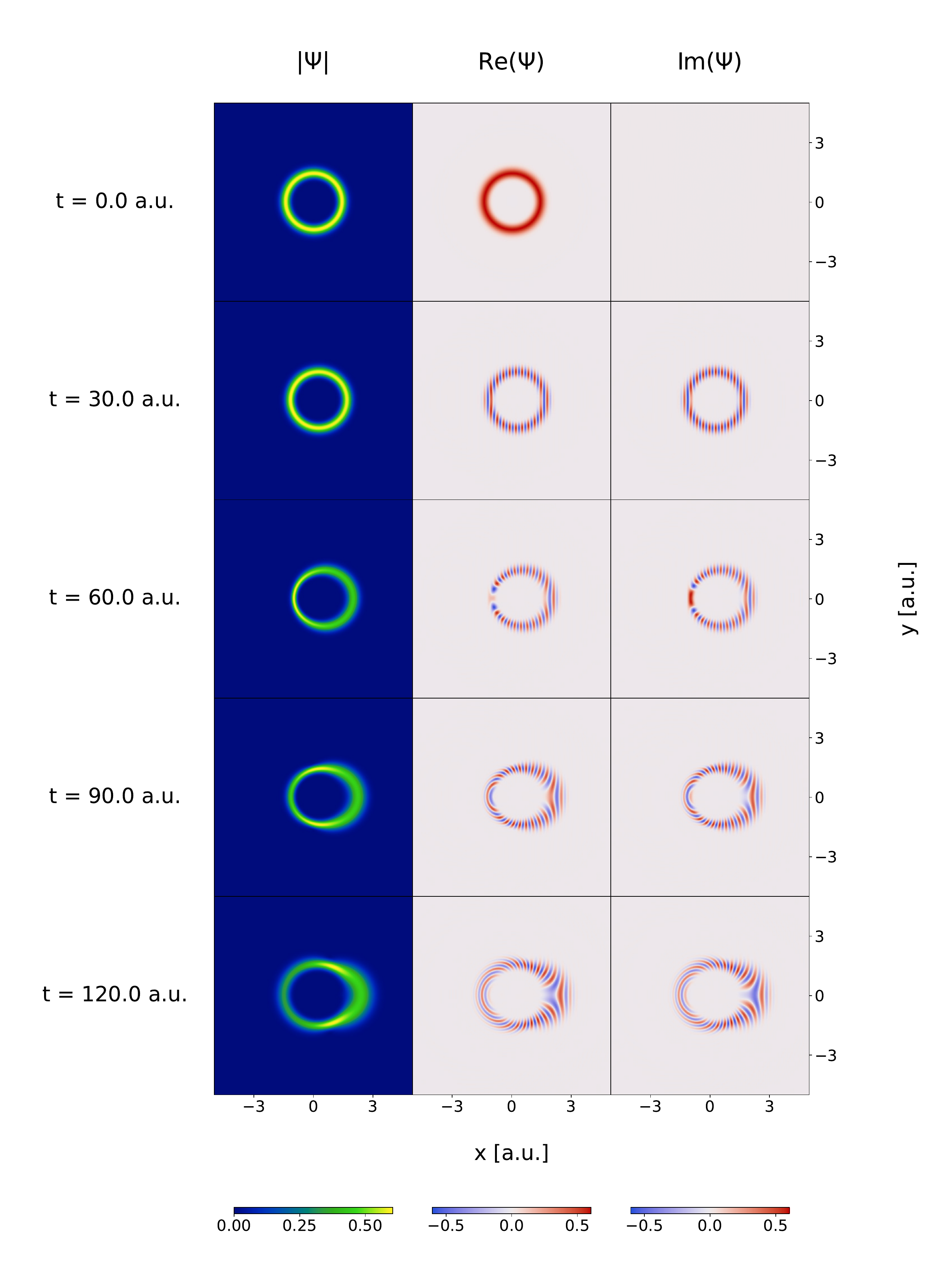}}
  \captionsetup{labelformat=empty}
  \caption{}
\end{figure}
\begin{figure}
  \ContinuedFloat
  \subfloat{\includegraphics[width=0.9\textwidth]{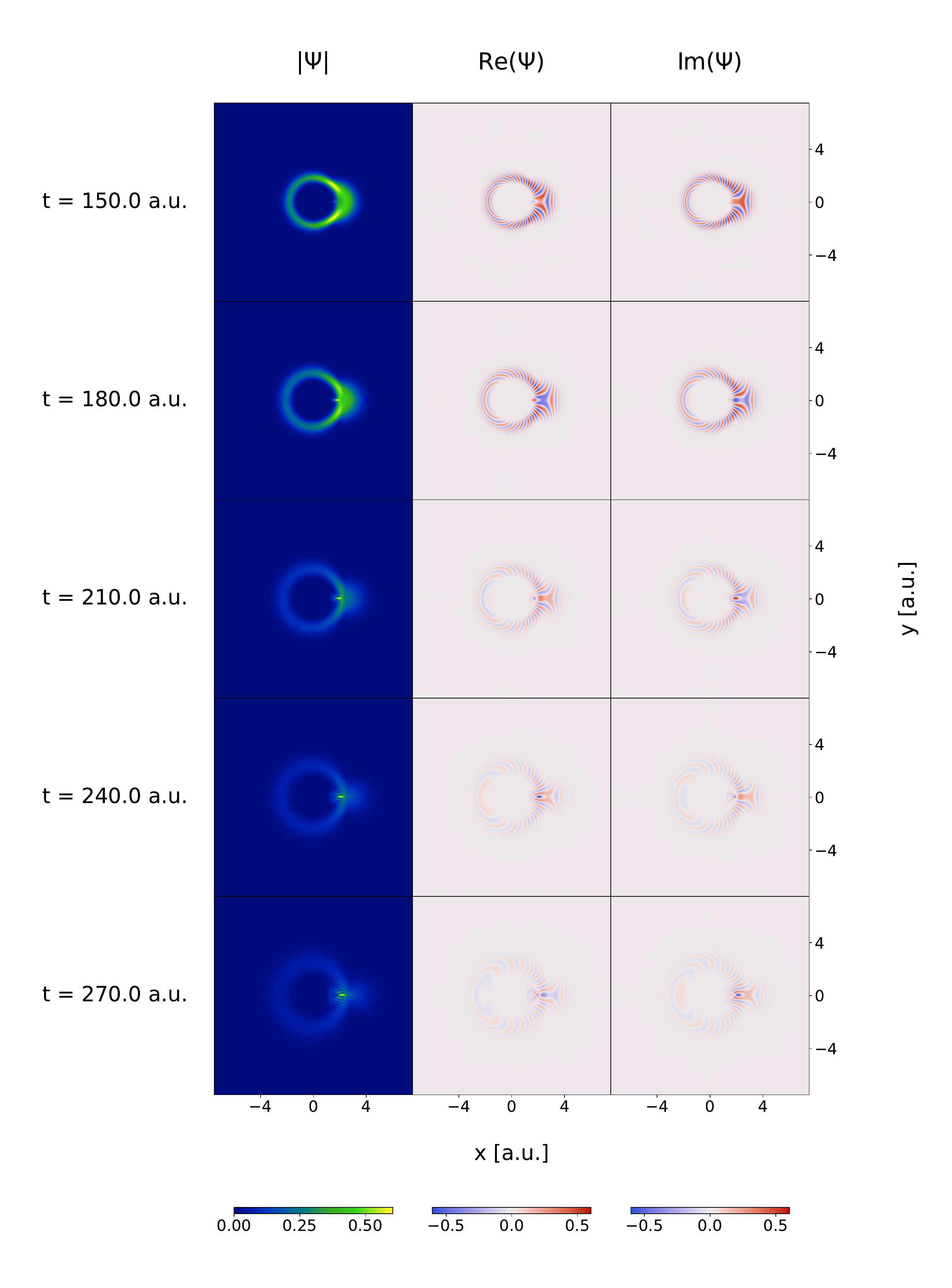}}
  \caption{ Same as in Fig.~\ref{fig:hydrogen_grid} but for the Morse model in the time interval from $t=0.0$ a.u. to $t=270.0$ a.u.}
  \label{fig:morse_grid}
\end{figure}

Fitting of the wave functions obtained in 
the Coulomb and Morse
simulations with ECGs is described in the next section.
Based on the analysing of the simulation results,
it is clear that Gaussians used in the fitting
need to be capable of describing the ground-state
wave function, the cylindrical deformation 
and oscillation of the function
due to the interaction with the field, the diffusion of
the function associated with 
possible ionization or/and
dissociation of the system, and the coupling
of the motion of all particles forming the system
including the non-BO
coupling of the motions of the electrons and 
the nuclei. Most of these features 
appear in the calculations of molecular static ground and 
excited states and an excellent performance of the ECGs
have been well documented in those calculations.
The features, that do not appear in the static calculations
are ionization and dissociation. 
It necessitates that the wave function is
allowed to acquire some
plane-wave character. This can be achieved by
allowing the shift vector and
parameter matrix ${\mathbf A}$ in Gaussian \eqref{cecg}
to become complex. Such Gaussians, named by us the fully flexible explicitly correlated Gaussians (FFECGs), have the following
form:
\begin{equation}
\begin{aligned}
\phi \left(\mathbf{r}\right) = \exp \bigg\{ & - \Big[ \mathbf{r} - (\mathbf{q} + \mathrm{i}\mathbf{p}) \Big]^{\prime} \Big({\mathbf A} + \mathrm{i}\mathbf{B} \Big) \\
& \Big[ \mathbf{r} - (\mathbf{q} + \mathrm{i}\mathbf{p} ) \Big] \bigg\},
\end{aligned}
\end{equation}
or alternatively the following form:
\begin{equation}
\begin{aligned}
\phi({\mathbf r}) = \exp \bigg[ - & \Big( \mathbf{r}-\mathbf{q} \Big)^{\prime} \Big( \mathbf{A} + {\rm i} {\mathbf B} \Big) \Big( {\mathbf r} -\mathbf{q} \Big) \\
& + \mathrm{i} \mathbf{p}^{\prime} \Big( \mathbf{r}-\mathbf{q} \Big) + \Big( \xi + \mathrm{i} \zeta \Big) \bigg],
\end{aligned}
\end{equation}
where ${\mathbf A}$ and ${\mathbf B}$, as defined before,
are real symmetric $3n\times3n$ matrices, and 
$\mathbf{p}$ and $\mathbf{q}$ are $3n$ vectors.
The FFECGs are 
fully flexible complex multi-particle Gaussians
that can provide a basis set for
expanding a time-evolving 
non-BO wave packet of a molecular system interacting 
with a short fast-varying intense laser pulse.
As shown in the next section, 
a linear combination of FFECGs can be used  
to very accurately represent the ground-state
wave functions of the two models considered 
in this work. It can also describe the time-dependent
oscillations of the time-evolving wave packet.
Also, due to making the Gaussian shift vectors
to be complex, the ionization and dissociation 
processes can be described.
And finally, FFECGs can very effectively represent
the coupled and highly correlated motions
of the electrons and nuclei forming the system.
They allow 
the electrons and the nuclei to be treated on an
equal footing in the calculation.

\section{Fitting the grid-based wave packet with FFECGs}  \label{fitting}

The 2D wave packet obtained in the time-dependent
grid calculation at each particular time step for each of the two models 
is fitted with a linear combination of FFECGs.
In this case FFECG has the following form:
\begin{equation}
\begin{split}
\phi({\mathbf r}) = \exp \Bigg[ & - (x, y) \left(
\begin{array}{cc}
a + \mathrm{i} b & 0 \\
0 & a + \mathrm{i} b 
\end{array}
\right) (x, y)^{\prime} \\
& + \mathrm{i} (p_x, p_y)(x,y)^{\prime} + 
(q_x, q_y)(x,y)^{\prime} \Bigg], \nonumber
\end{split}
\end{equation}
where ${\mathbf p} = (p_x, p_y)^{\prime}$ and
${\mathbf q} = (q_x, q_y)^{\prime}$.
Thus, in the 2D case $\phi({\mathbf r})$,
is dependent on six real non-linear parameters,
$a$, $b$, $p_x$, $p_y$, $q_x$, and $q_y$.
As the time-dependent calculations for both models are
initiated with the corresponding ground-state
wave functions, which are real and spherically 
symmetric, appropriate 
FFECGs need to be used. 
For the hydrogen model, these FFECGs
are simple spherical Gaussians centered at $x=y=0$
with $a \ne 0$ and $b = p_x = p_y = q_x = q_y = 0$. 
A linear combination of FFECGs
of this kind can fit the ground-state wave function
obtained in the grid calculation with very good accuracy.
For the Morse model, one needs to only use FFECGs 
with $a \ne 0$, $b \ne 0$, and $p_x = p_y = q_x = q_y = 0$
to generate a good fit to the ground-state 
wave function obtained in the grid calculation.
Additionally, for each FFECG pair used, the 
two Gaussians need to have the same $a$, but their $b$'s
should be $b$ and $-b$, so they can be "contracted"
for form the following real function:
\begin{align}
\frac{\phi(-b) - \phi(+b)}{2\mathrm{i}} &= K \exp(-a (x^2 + y^2) \sin(b (x^2 + y^2))) = \nonumber \\
&= \exp(-ar^2) \sin(b r^2),
\end{align}
where $K$ is a constant.
A linear combination of such contracted FFECGs
provides a very accurate fit to the ground-state 
wave function for the Morse model.

The fitting of the grid ground-state 
wave functions with FFECGs
are done using the standard least-squares implementation from the SciPy Python library\cite{SciPy}.
The method is also used to fit the wave functions
obtained in consecutive time steps obtained in the grid
simulation. First, the number of FFECGs used
is the same as for the ground state, but the 
parameters frozen at zero for the 
ground-state wave function are now unfrozen and 
optimized.
For both models we set the threshold for the least squares cost function equal to $10^{-5}$.
When the assumed accuracy of the fitting
cannot be achieved, additional FFECGs are added to the
basis set. Each addition includes a group of FEECGs with parameters:
\begin{align}
&\phi(a=1, b=0, p_x = 1, p_y = 0, q_x = 1, q_y = 0), \nonumber \\
&\phi(a=1, b=0, p_x = 1, p_y = 0, q_x = -1, q_y = 0), \nonumber \\
&\phi(a=1, b=0, p_x = -1, p_y = 0, q_x = 1, q_y = 0), \; \mathrm{and} \nonumber \\
&\phi(a=1, b=0, p_x = -1, p_y = 0, q_x = -1, q_y = 0). \nonumber
\end{align}
The fitting that involves optimization of all
linear and nonlinear parameters of the enlarged 
FFECG basis set continues for some number 
of the following time steps until it is determined
that the fitting process is no longer successful.
At that point new FFECGs are added to the basis set
and the wave packet is refitted.
The fitting for the Coulomb and Morse models
is shown in 
Figs.~\ref{fig:hydrogen_gaussian} and \ref{fig:morse_gaussian}.
In the figures, the wave packets obtained in the
grid calculations are compared with the corresponding 
FFECG fits for some selected time points.

\begin{figure}
  \centering
  \subfloat{\includegraphics[width=0.85\textwidth]{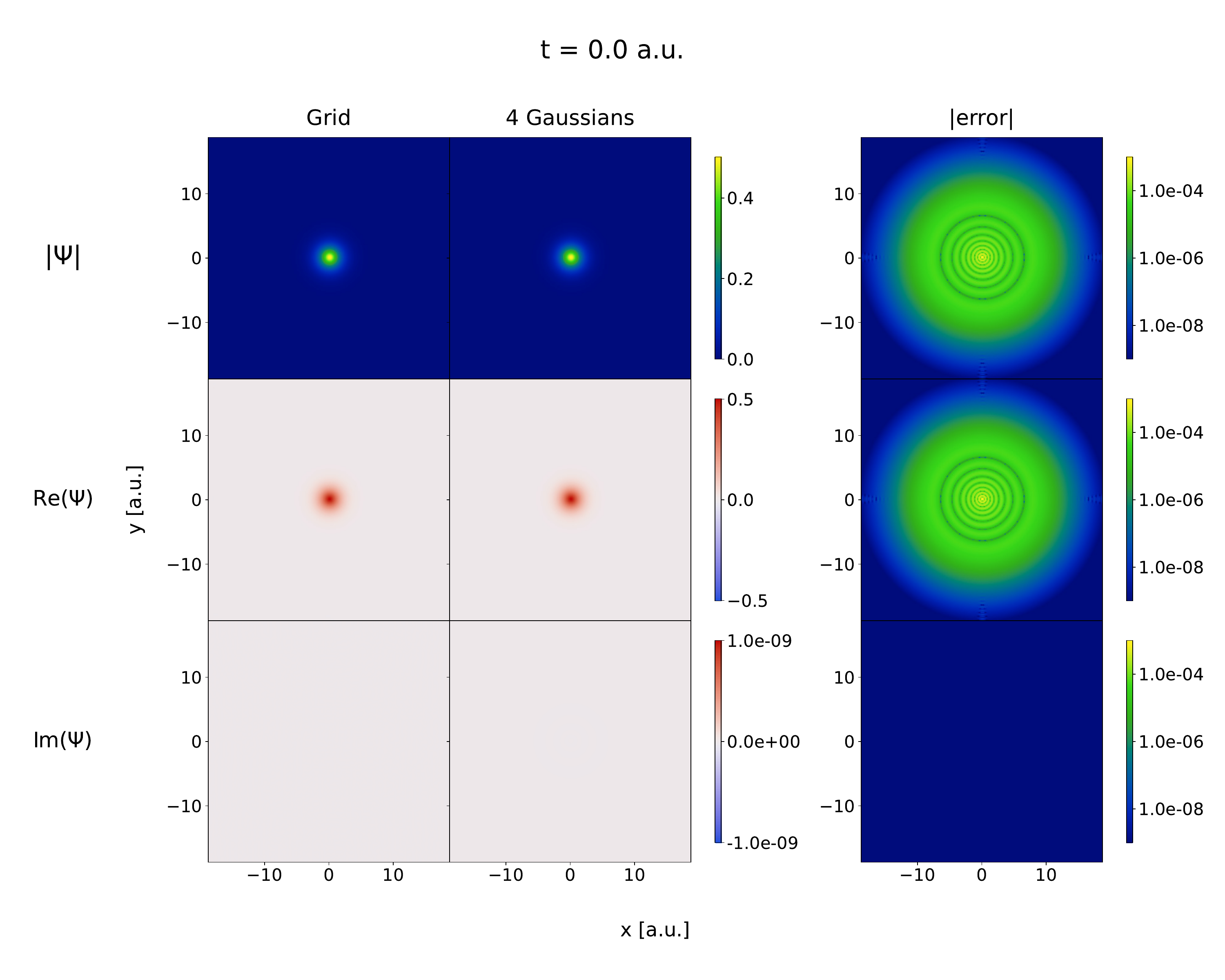}}

  \subfloat{\includegraphics[width=0.85\textwidth]{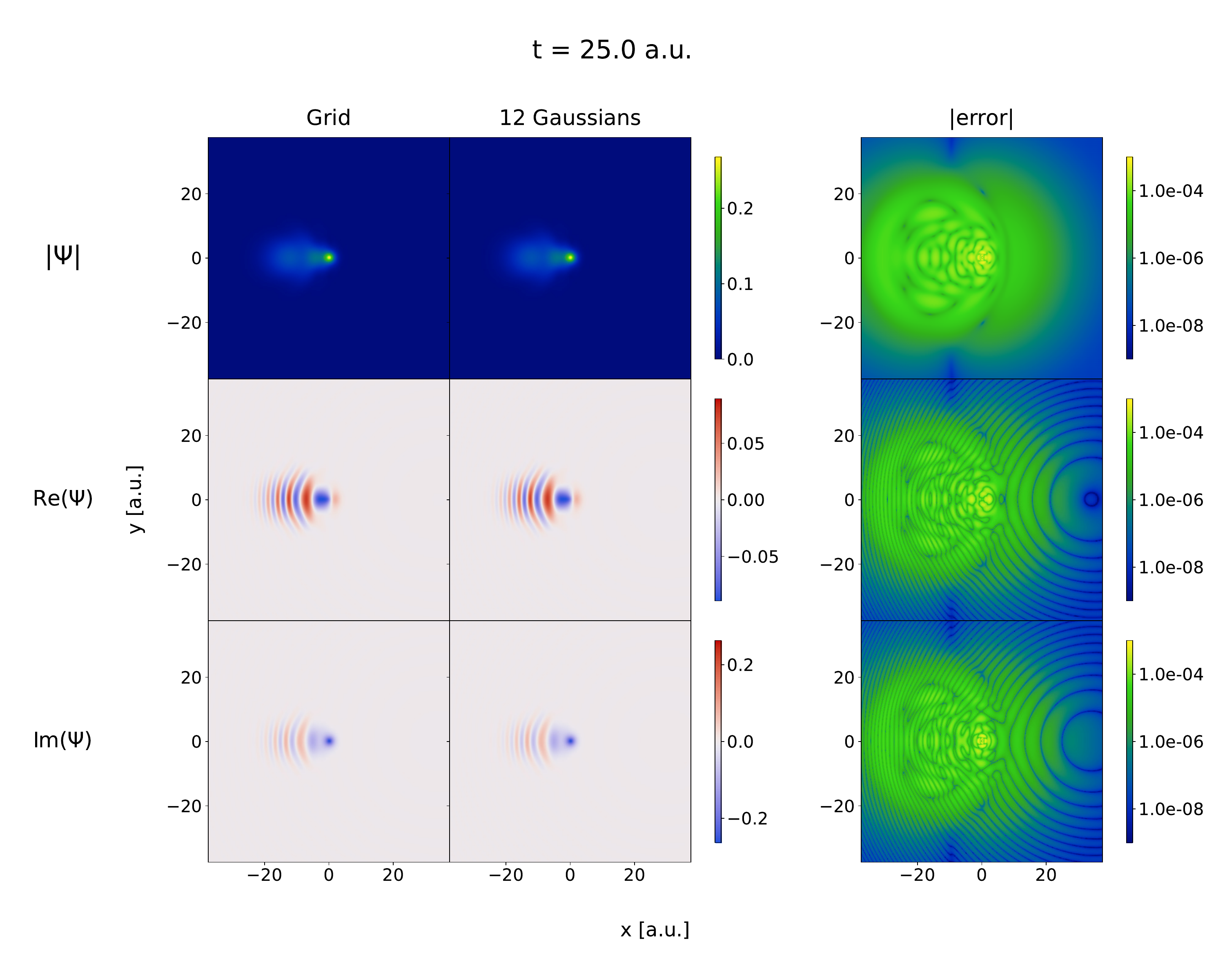}}
  \captionsetup{labelformat=empty}
  \caption{}
\end{figure}
\begin{figure}
  \ContinuedFloat
  \centering
  \subfloat{\includegraphics[width=0.85\textwidth]{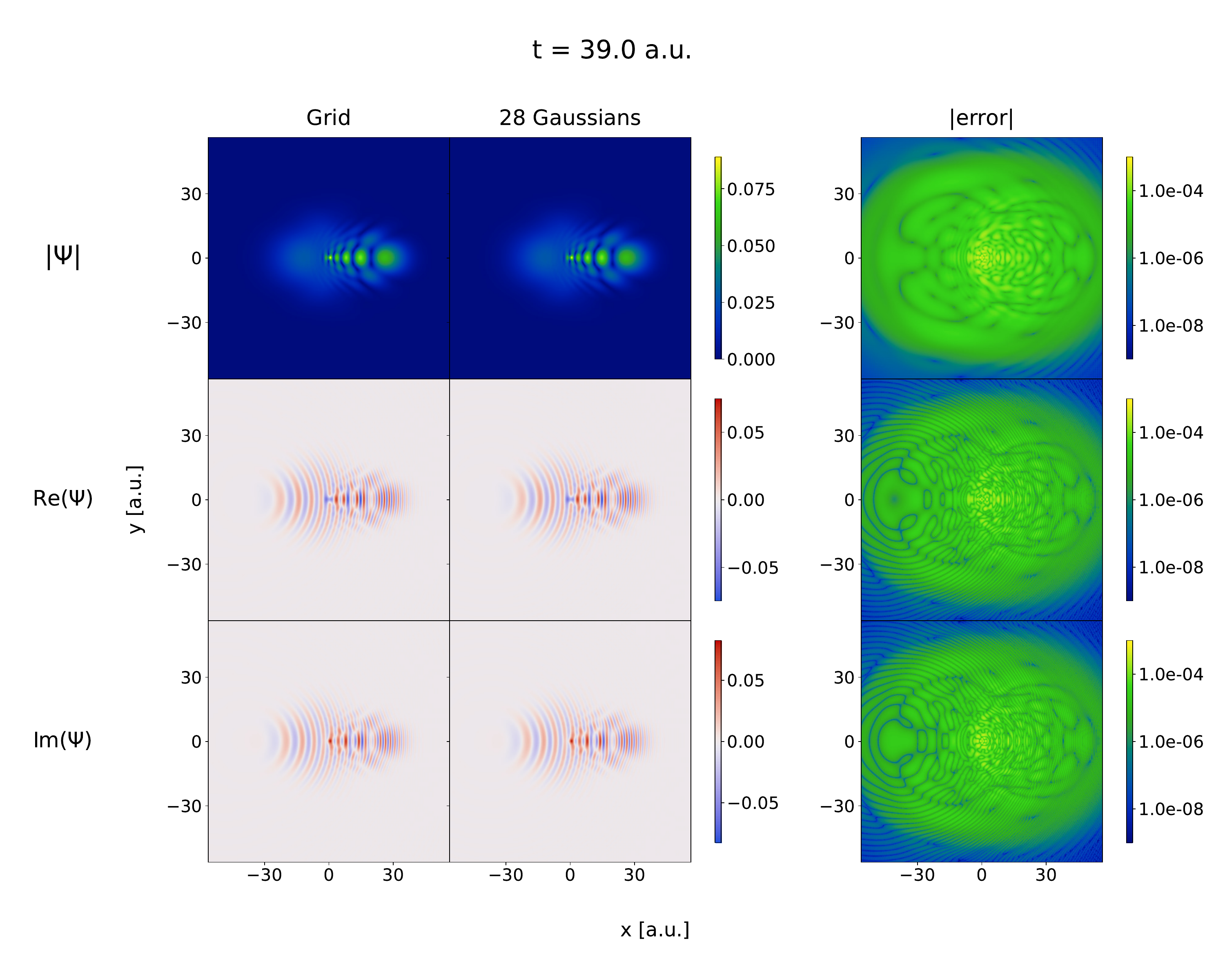}}

  \subfloat{\includegraphics[width=0.85\textwidth]{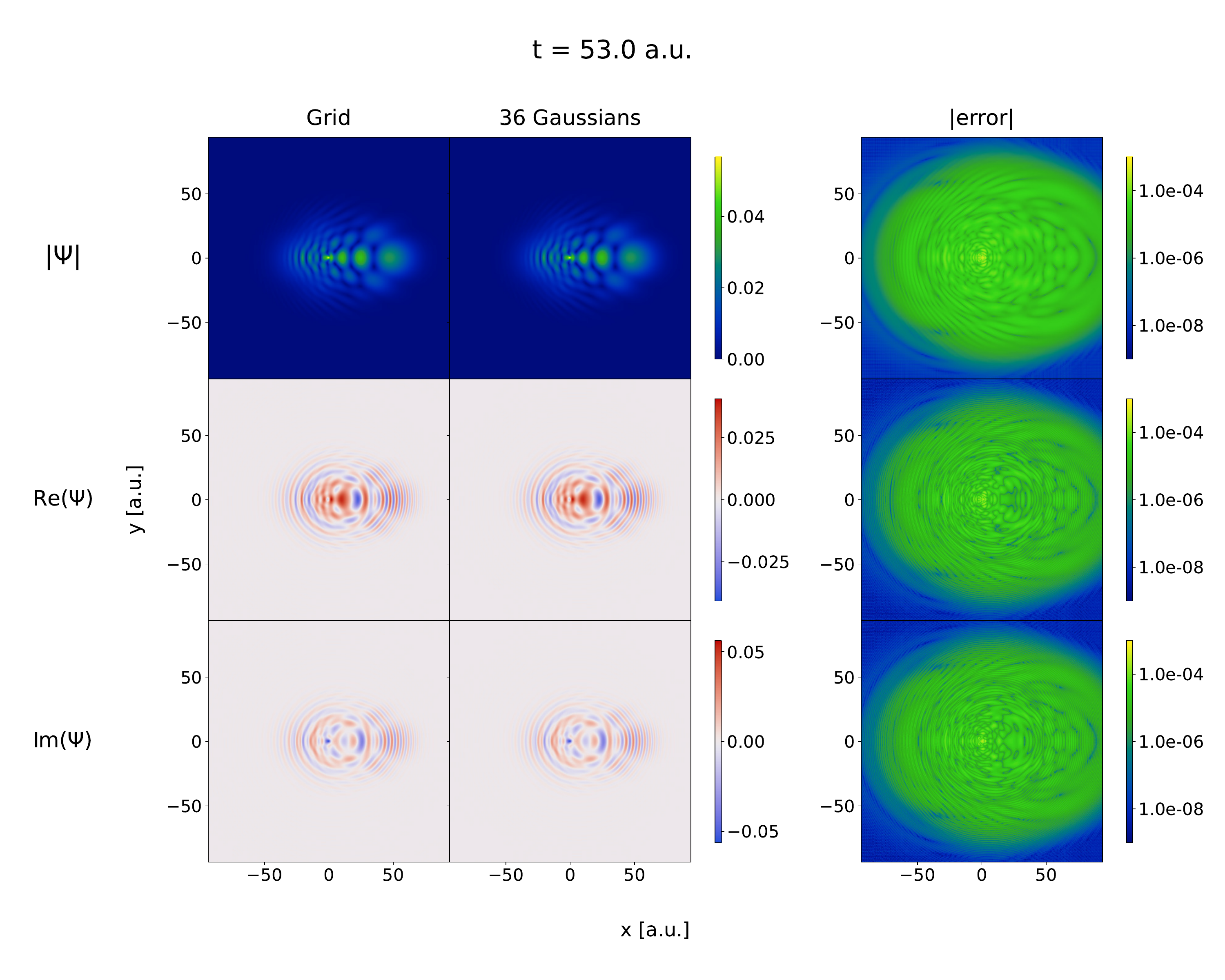}}
  \caption{FFECG fits to the selected time frames extracted from the grid-based simulation trajectory of the Coulomb model. The error represents the difference in either the absolute value, the real part, or the imaginary part between the grid wave function and the optimized linear combination of Gaussian functions }
  \label{fig:hydrogen_gaussian}
\end{figure}

\begin{figure}
  \centering
  \subfloat{\includegraphics[width=0.85\textwidth]{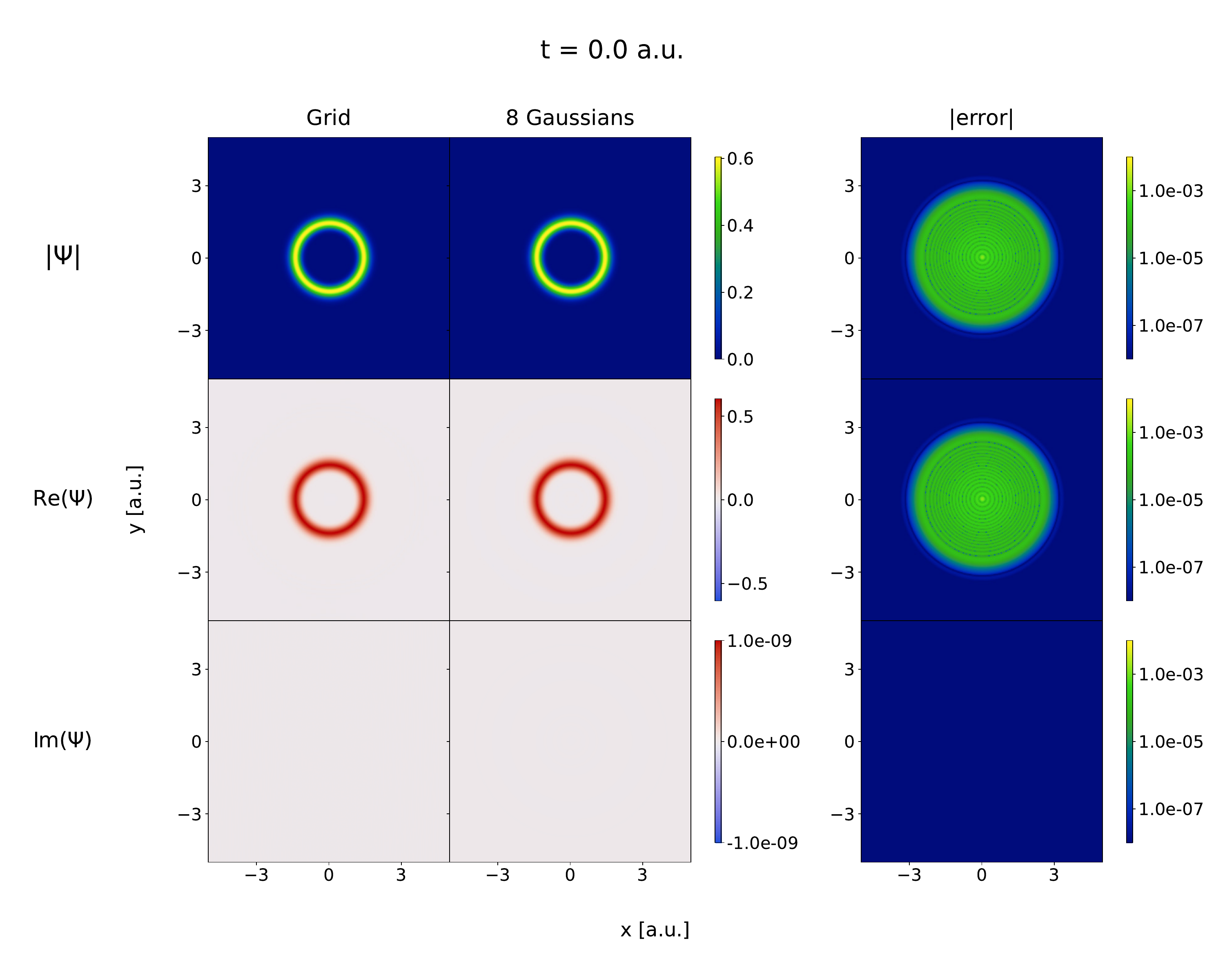}}
  
  \subfloat{\includegraphics[width=0.85\textwidth]{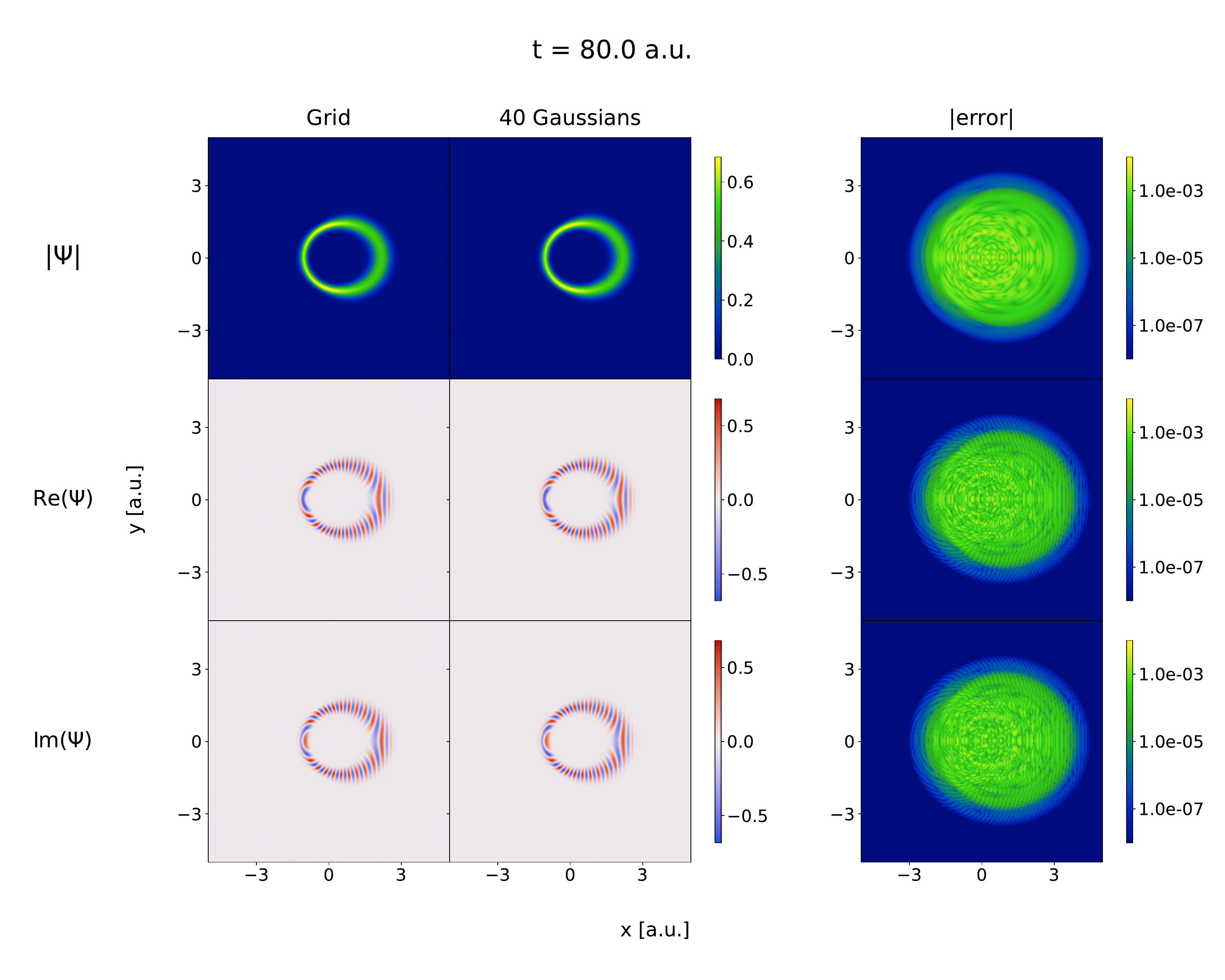}}
  \captionsetup{labelformat=empty}
  \caption{}
\end{figure}
\begin{figure}
  \ContinuedFloat
  \centering
  \subfloat{\includegraphics[width=0.85\textwidth]{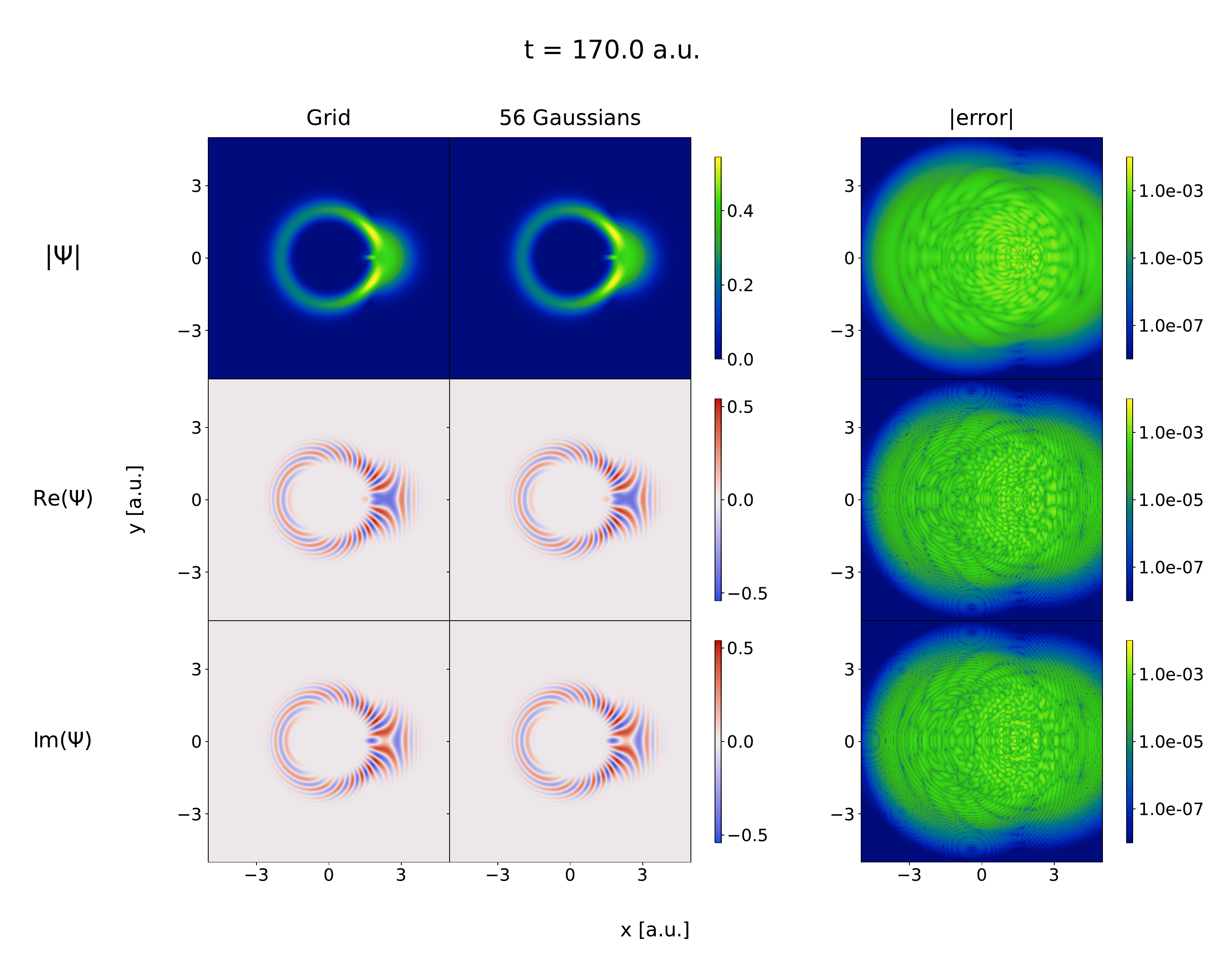}}

  \subfloat{\includegraphics[width=0.85\textwidth]{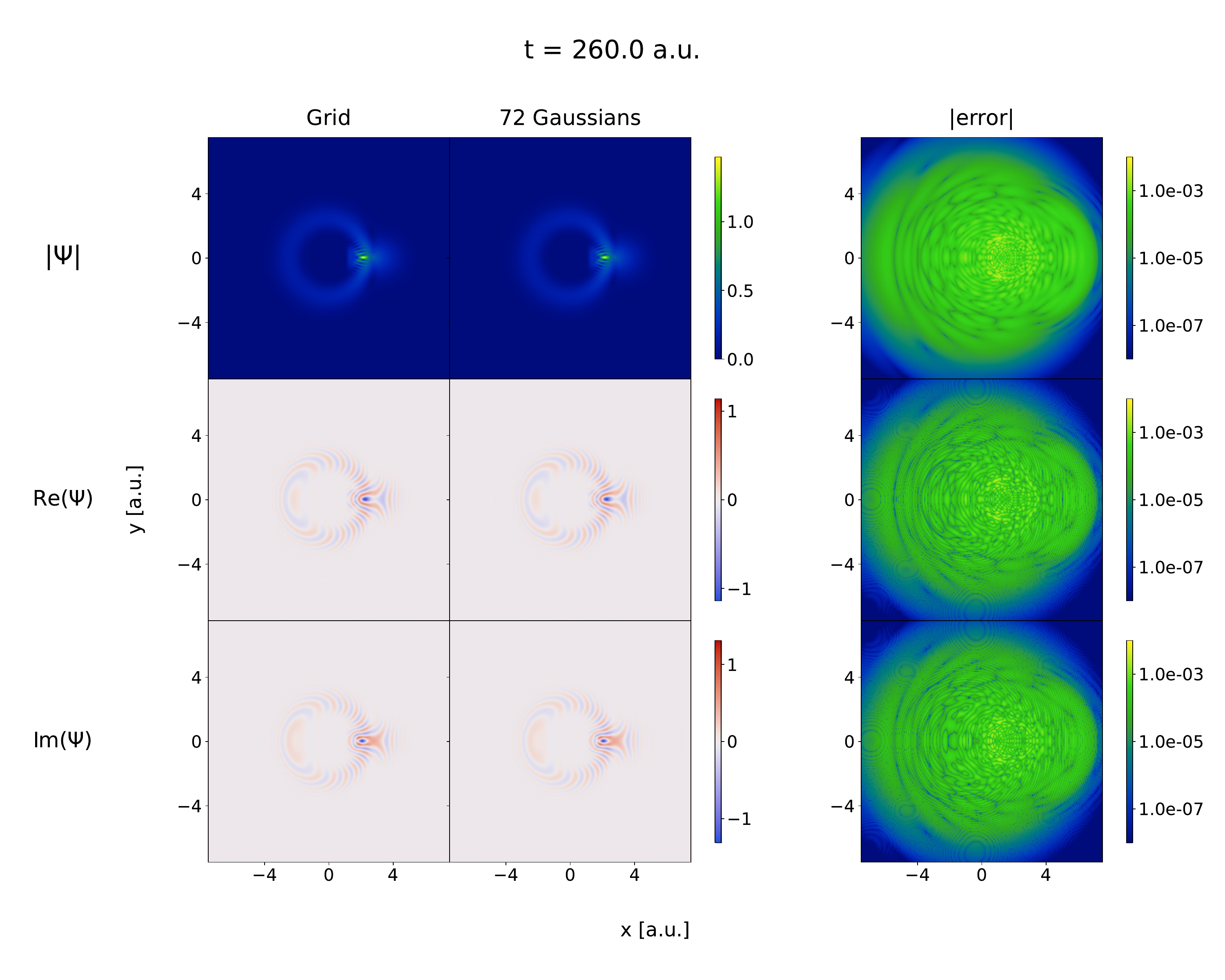}}
  \caption{ Same as Fig.~\ref{fig:hydrogen_gaussian} but for the Morse model }
  \label{fig:morse_gaussian}
\end{figure}

\begin{figure}
  \centering
  \includegraphics[width=0.49\linewidth]{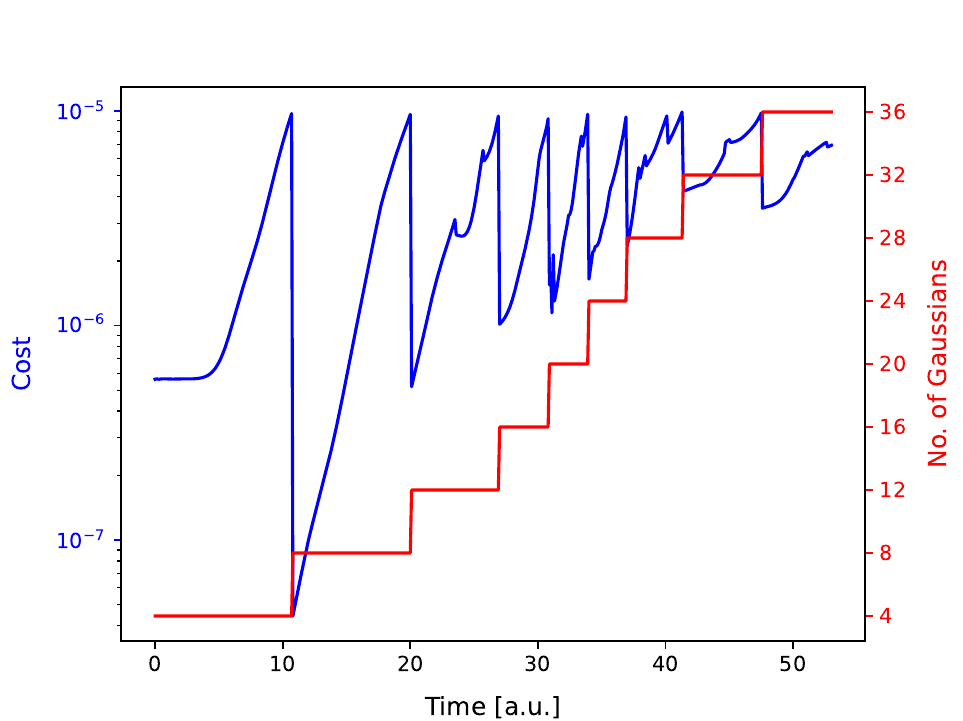}
  \includegraphics[width=0.49\linewidth]{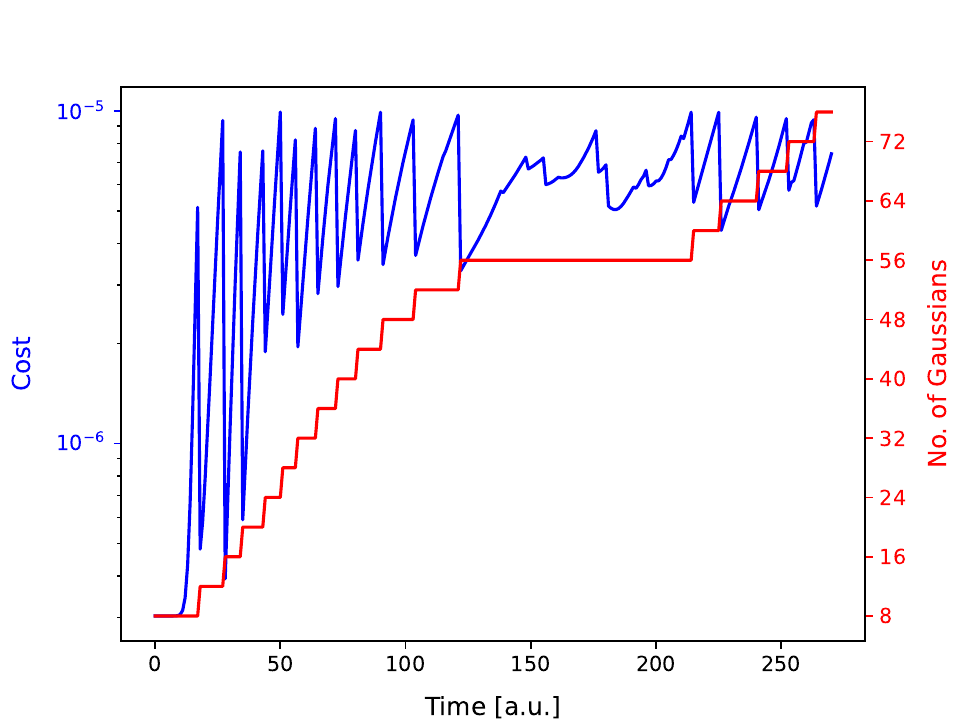}
  \caption{ Evolution of the cost function and FFECG basis set size over time during the fitting process for the Coulomb model (left) and for the Morse model (right). A new set of functions is added to the total basis set whenever the cost function reaches the predetermined threshold of $10^{-5}$, ensuring the maintenance of the desired level of accuracy.}
  \label{fig:cost}
\end{figure}

\begin{figure}
  \centering
  \includegraphics[width=0.99\textwidth]{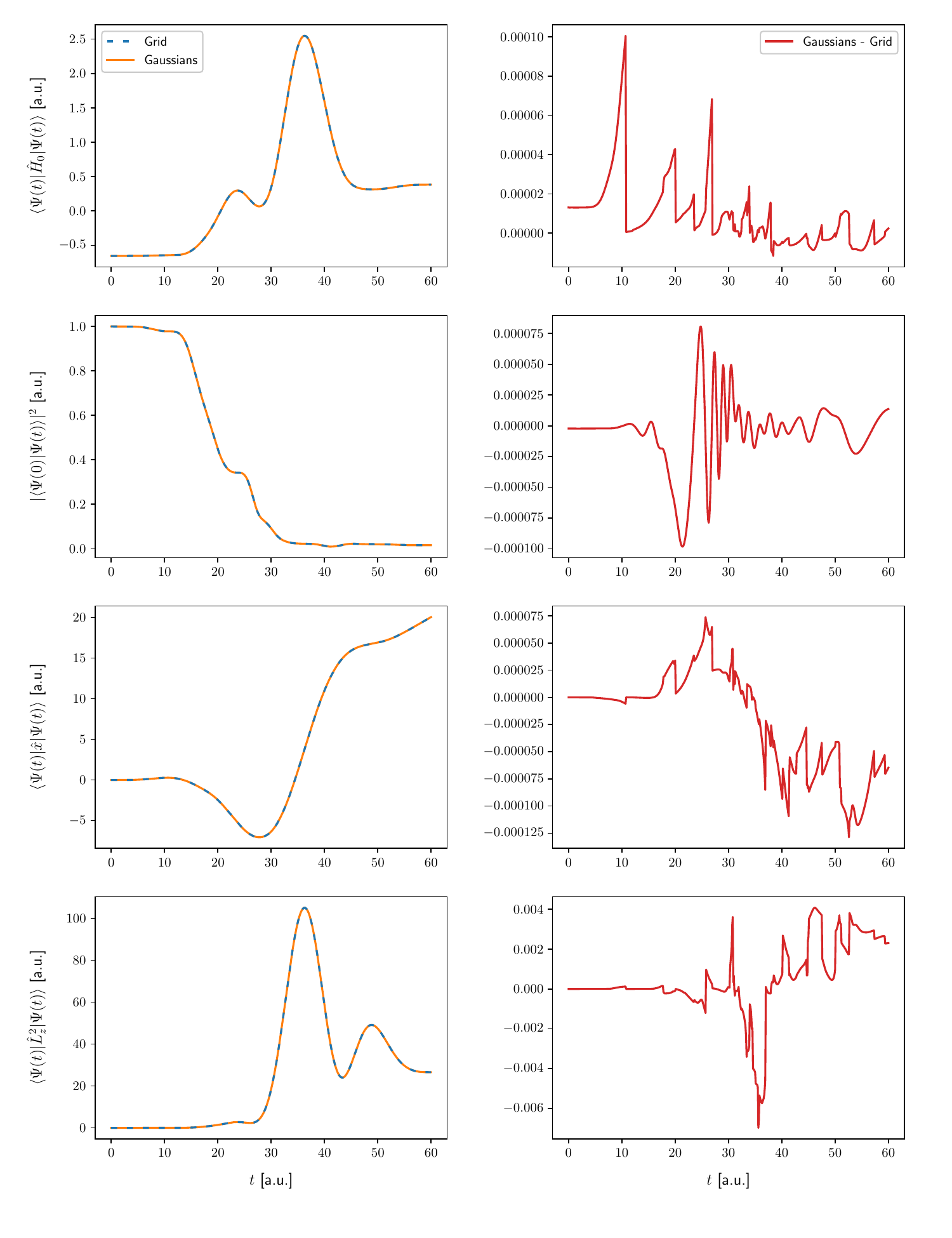}%
  \caption{ Left: Time-resolved observables computed for the Coulomb model, with the reference grid wave function (dashed lines), and with the fitted Gaussian wave function (solid lines). Right: Differences between grid and Gaussian observables. From top to bottom: expectation value of the field-free Hamiltonian, projection of the wave packet onto the ground-state wave function (the initial state), dipole moment expectation value, expectation value of the squared $z$-component of the angular momentum.}
  \label{fig:hydrogen_obs}
\end{figure}

\begin{figure}
  \centering
  \includegraphics[width=0.99\textwidth]{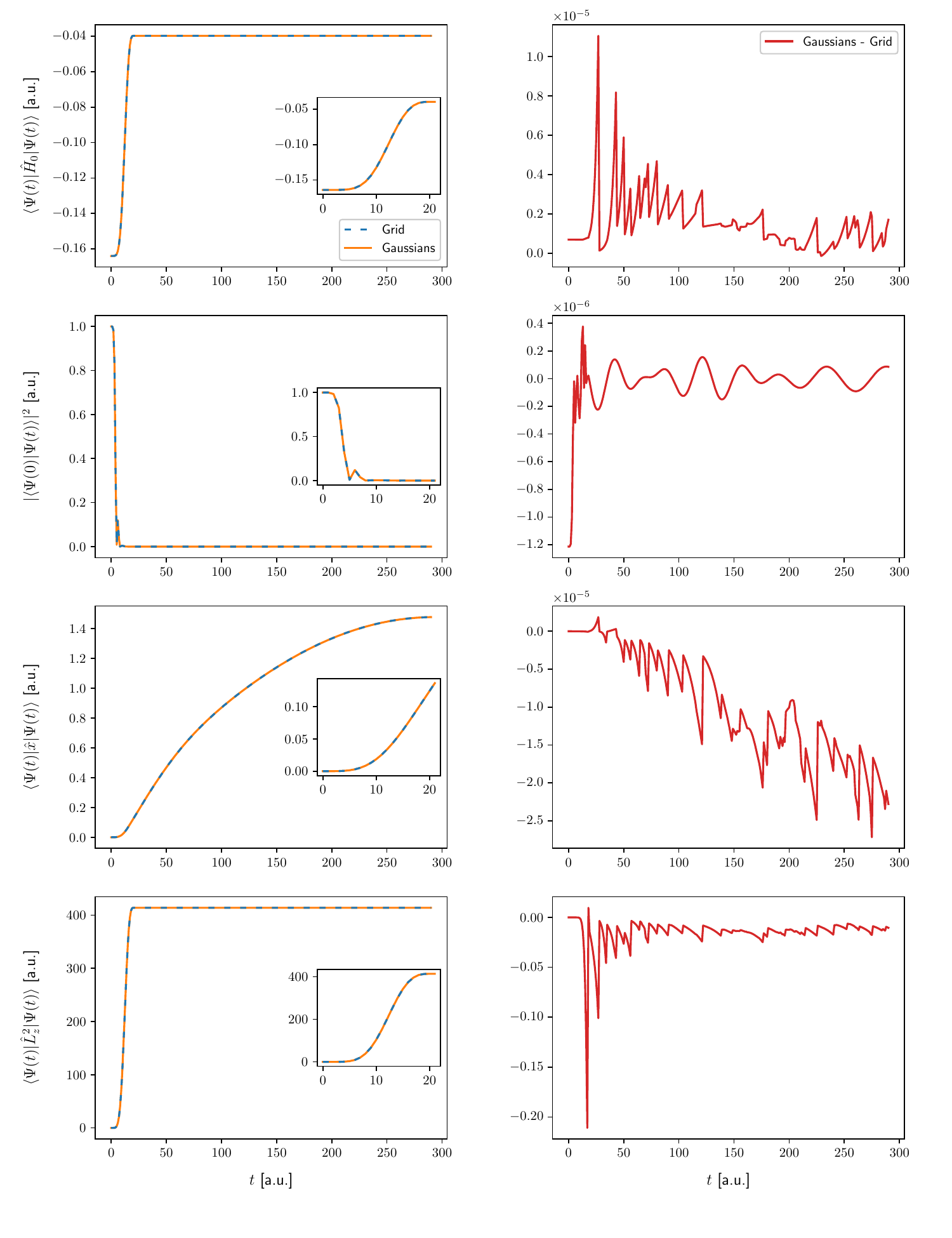}%
  \caption{ Same as Fig.~\ref{fig:hydrogen_obs} but for the Morse model. The insets display the evolution of the observables over the duration of the laser pulse (0-20 a.u).}
  \label{fig:morse_obs}
\end{figure}

Also, in Fig.~\ref{fig:cost}
the changes of the cost
function representing the accuracy of the fitting, as well as the time-resolved size of the basis set,
are plotted for the Coulomb and Morse 
models.
Both Figs.~\ref{fig:hydrogen_gaussian} and \ref{fig:morse_gaussian}, and Fig.~\ref{fig:cost}
show that the
fitting accuracy is satisfactory for both models
despite the wave function becoming increasingly 
more complicated. Naturally, as this happens, the 
number of FFECGs has to be constantly increased.

In general, the Morse model seems to be somewhat more difficult to represent using FFECGs than the Coulomb model. The reason for this behavior can be attributed to the maximum of the density of the second nucleus in the Morse model being shifted away from the reference nucleus by some distance (in the ground-state this distance is approximately equal to the equilibrium internuclear distance). This type of shifting does not happen in the Coulomb model. The shifting of the density maximum away from the reference nucleus required including more ECGs in the wave function. However, if this is done, the ECG expansion of the wave packet in the Morse model should be equally accurate as it is for the Coulomb model.

The least-square fitting of a linear combination of ECGs to a grid wave function produces a wave packet that provides a representation of the grid function with uniform spatial quality. ECG representations obtained by minimization of energy-based functionals (e.g. the variational Rayleigh-Ritz functional or Rothe variational functional) usually are more accurate in some spatial domains than the others. It is difficult to {\it a priori} determine which imperfections of the representation of the wave function will be amplified and which will be suppressed when a particular observable is calculated.
 
To better elucidate the accuracy of the FFECG fits
obtained in this work, we compare
time-resolved 
observables obtained for the grid wave packet
and the fitted wave packet.
The calculated observables are the expectation value of the field-free Hamiltonian, 
the ground-state survival probability (i.e., the square of
the autocorrelation function),
the dipole moment expectation value, 
and the expectation value of the squared 
$z$-component of the angular momentum.
The comparison is shown in Fig.~\ref{fig:hydrogen_obs}
for the Coulomb model and in Fig.~\ref{fig:morse_obs}
for the Morse model.
On the left-hand side of each figure the time-resolved
plots of each of the four observables for
the grid and FFECG wave packets are shown together in
four separate frames. As one can see, within the 
left-hand-side plots, the curves corresponding
to the two wave packets for all four observables for
both models are practically indistinguishable.
The right-hand-side plots do indicate some error
fluctuations, but these are of similar magnitude
as the fluctuations in the fitting error and, therefore,
acceptable.

We note that the ground-state survival probability of the final
state is small or zero for both models, and that the angular-momentum
expectation value indicates involvement of highly excited rotational
states. The final energy
for the Coulomb model is positive, indicating an unbound (i.e., ionized) state.
For both models, the increase of the dipole moment during the dynamics indicate
the large, asymmetric spreading of the wave packet.
It is quite remarkable that so relatively few FFECGs are required to accurately
reproduce such complicated dynamics.

\section{Conclusion} \label{conclusion}

The grid approach is used to obtain solution 
of the time-dependent Schr\"{o}dinger equation
for two 2D model systems that represent features
which appear in quantum-dynamics
time-propagation of the wave packet
representing a diatomic neutral molecule interacting
with a short intense laser pulse and performed
without assuming the Born-Oppenheimer approximation.
The grid wave functions obtained in consecutive
time steps are fitted with a combination Gaussian 
functions that are 2D versions of more the general 
fully-flexible explicitly correlated Gaussians 
(FFECGs) with complex exponential
parameters and complex shifts of the Gaussian centers.
The fitting procedure employs the least-square procedure
and involves growing the basis set of the Gaussians
to provide a uniformly good fit for a representative set
of time points obtained from the grid time propagation.
The two models considered in the calculations 
involve a single-particle in the central potential 
represented by an attractive Coulomb interaction and
a Morse potential. Based on the results obtained in 
the calculations we can expect that FFECGs will
provide a good basis set for laser-induced non-BO dynamics
of a diatomic molecule. Work on implementing 
FFECGs in molecular QD simulations is in progress.

Finally, this work represents a preliminary step in the application of FFECGs to describe the coupled electronic-nuclear dynamics in atomic and molecular systems.
In the future work involving FFECGs and the non-BO nuclear-electronic quantum dynamics, the Rothe method\cite{rothe1930,bornemann1990,horenko2004} will be employed to propagate the wave packet. The approach, also known as the adaptive method of time layers\cite{Deuflhard2012_}, relies on reformulating the time-dependent variational principle into a series of minimizations of the Rothe functional at consecutive time steps. This is an alternative to the standard real-time propagation techniques based on the Dirac--Frenkel variational principle and propagated using, e.g., Runge--Kutta methods. The ECG optimization protocol developed in the present work to fit a linear combination of ECG to the grid-based wave packet will be applied to minimize the Rothe functional with respect to the linear and non-linear parameters of the ECGs. In our works on the variational calculations of molecular stationary states with real and complex ECGs we have developed procedures for the variational minimization of the energy functional which employs the analytical energy gradient determined with respect to the ECG nonlinear parameters. The use of the gradient has significantly expedited the functional minimization and enabled to obtain non-BO energies and the corresponding wave functions whose accuracy by far exceeds the results obtained by others. The gradient-based approach will also be used in the optimization of the ECGs parameters carried out through the minimization of the Rothe functional. The high efficiency of the computer code for the optimization of the ECGs in the stationary-state calculations has been also achieved by deriving the algorithms for calculating the necessary N-particle matrix elements (i.e. the overlap, Hamiltonian, and gradient matrix elements) using the matrix differential calculus and by coding them using highly parallel, vectorized, and GPU-enabled strategies. These strategies will be used in the Rothe time-propagation calculations.

\begin{acknowledgement}
    This work was supported by the Research Council of Norway through its Centres of Excellence scheme, project no. 262695.
    Partial support from the National Science Foundation (grant no. 1856702) is also acknowledged.
    A. P. W. also acknowledges support from Polish National Science Centre (NCN) through Grant No. 2017/25/B/ST4/02698.
    The calculations presented in this work were carried out using resources provided by University of Arizona Research Computing and by Wroclaw Centre for Networking and Supercomputing, Grant No. 567.
\end{acknowledgement}



\bibliography{bibliography}

\end{document}